\documentclass[a4paper,UKenglish,cleveref,pdfa]{lipics-v2021}
\nolinenumbers
\ArticleNo{1}
\hideLIPIcs
\usepackage{hyperref}
\usepackage{booktabs} 
\usepackage{amssymb,amsmath,amsthm,mathtools}
\usepackage[inline]{enumitem}
\usepackage{tikz}
\usetikzlibrary{calc,positioning}
\usepackage{tikz-cd}
\usepackage{xcolor}
\usepackage{hyperref}
\usepackage{comment}
\usepackage{calc}
\usepackage{longtable}

\newcommand{\TA}{{\mathsf{TA}}}

\newcommand{\Sig}{\mathsf{Sig}}
\newcommand{\Mod}{\mathsf{Mod}}
\newcommand{\Sen}{\mathsf{Sen}}

\newcommand{\Set}{\mathbb{S}\mathsf{et}}
\newcommand{\A}{\mathfrak{A}}
\newcommand{\B}{\mathfrak{B}}
\newcommand{\D}{\mathfrak{D}}

\newcommand{\card}{\mathsf{card}}
\newcommand{\act}{\mathfrak{a}}
\newcommand{\Space}{~~~~~}
\newcommand{\pos}[1]{{\langle}#1{\rangle}}
\newcommand{\Forall}[1]{\forall #1\,{\cdot}\,}
\newcommand{\Exists}[1]{\exists #1\,{\cdot}\,}
\newcommand{\proofrule}[2]{\displaystyle\frac{#1}{#2}}
\newcommand{\red}{\mathord{\upharpoonright}}
\usepackage{scalerel}

\newcommand{\bbsemicolon}{%
  \scalerel*{%
    \hbox{\usefont{U}{bbold}{m}{n} ;}%
  }{;}%
}
\newcommand{\comp}{\mathbin{\bbsemicolon}}
\newcommand{\minisec}[1]{%
  \par\addvspace{\smallskipamount}\noindent%
  \textit{#1}.\enspace%
}
\newenvironment{grammar}{%
  \def\ceqdef{\mkern 6mu\Coloneqq\mkern 6mu}%
  \def\alt{\mkern 4mu\mid\mkern 4mu}%
  \[%
}{\]\ignorespacesafterend}
\setlist[enumerate,1]{
  font=\normalfont,
  label={\arabic*.},
  ref=\arabic*
}

\newlist{inlinenum}{enumerate*}{1}
\setlist[inlinenum,1]{
  font=\normalfont,
  label=(\emph{\alph*})
}

\setlist[itemize,1]{
}

\setlist[description]{
  font=\normalfont\em,
  leftmargin=\parindent
}

\newlist{plainlist}{itemize}{1}
\setlist[plainlist]{
  label={},
  leftmargin=0pt,
  itemsep=\parskip
}
\newcommand*{\pcformat}[1]{%
  [\;{\normalfont\itshape #1}\;]%
}

\newenvironment{proofcases}[1][]{%
  \description[font=\pcformat, leftmargin=\parindent, #1]%
}{\enddescription}

\usepackage{array,etoolbox}
\makeatletter

\newlength{\PS@lastparam}
\newlength{\PSlastparam}
\newcommand{\PSlp}{%
  \setlength{\PSlastparam}{\PS@lastparam}%
  \the\PSlastparam
}
\def\PS@sub@lastparam{}

\newcommand{\PS@numwidth}{99}
\newcommand{\PSnumwidth}[1]{%
  \renewcommand{\PS@numwidth}{#1}%
}

\newcommand{\PS@style}{\small}
\newcommand{\PS@numstyle}{\footnotesize}

\newlength{\PSindent}
\newlength{\PS@extraindent}
\newlength{\PSpre}
\newlength{\PSpost}
\newlength{\PS@Nwidth}
\newlength{\PS@Swidth}
\newlength{\PS@Ewidth}
\newlength{\PScolsep}
\newcommand{\PS@rownumber}{%
  \ifPS@subsubsteps
  \thePSsubstepc.%
  \the\numexpr\value{PSsubsubstepc}+1\relax
  \else
  \ifPS@substeps
  \thePSstepc.%
  \the\numexpr\value{PSsubstepc}+1\relax
  \else
  \the\numexpr\value{PSstepc}+1\relax
  \fi\fi
}
\newcommand{\PS@step}{%
  \ifPS@subsubsteps
  \refstepcounter{PSsubsubstepc}%
  \else
  \ifPS@substeps
  \refstepcounter{PSsubstepc}%
  \else
  \refstepcounter{PSstepc}
  \fi\fi%
}

\newif\ifPS@inprogress
\newif\ifPS@substeps
\newif\ifPS@subsubsteps
\newif\ifPS@continued
\newif\ifPS@subcontinued
\newcounter{PSc}
\newcounter{PSstepc}[PSc]
\newcounter{PSsubstepc}[PSstepc]
\renewcommand{\thePSsubstepc}{\thePSstepc.\arabic{PSsubstepc}}
\newcounter{PSsubsubstepc}[PSsubstepc]

\newenvironment{proofsteps}[1]{%
  \global\settowidth{\PS@lastparam}{\PS@style\hspace*{#1}}
  \ifPS@continued\else\refstepcounter{PSc}\fi
  \begingroup
  \setlength{\LTpre}{\PSpre}%
  \setlength{\LTpost}{\PSpost}%
  
  \setlength{\tabcolsep}{0pt}
  \noindent\PS@style
  \settowidth{\PS@Nwidth}{\PS@numstyle\PS@numwidth}%
  \setlength{\PS@Swidth}{#1}%
  \addtolength{\PS@Swidth}{-\PS@extraindent}%
  \setlength{\PS@Ewidth}{\linewidth}%
  \addtolength{\PS@Ewidth}{-\PSindent}%
  \addtolength{\PS@Ewidth}{-\PS@extraindent}%
  \addtolength{\PS@Ewidth}{-\PS@Nwidth}%
  \addtolength{\PS@Ewidth}{-\PScolsep}%
  \addtolength{\PS@Ewidth}{-\PS@Swidth}%
  \addtolength{\PS@Ewidth}{-\PScolsep}%
  \PS@inprogresstrue
  \longtable{%
    @{\hspace*{\PSindent}\hspace*{\PS@extraindent}\makebox[\PS@Nwidth][r]{\PS@rownumber}}%
    @{\hskip\PScolsep}>{\PS@step}p{\PS@Swidth}%
    @{\hskip\PScolsep}>{\footnotesize\raggedright\arraybackslash}p{\PS@Ewidth}%
  }%
}{%
  \ifPS@inprogress
  \addtocounter{table}{-1}%
  \endlongtable  
  \endgroup
  \PS@continuedfalse
  \PS@inprogressfalse
  \else\fi
}

\newcommand{\PSbreak}[1]{%
  \endproofsteps
  \par\medskip
  #1
  \medskip\par
  \PS@continuedtrue
  \proofsteps{\PS@lastparam}%
}

\newif\ifPS@sub@inprogress

\newif\ifPS@laststep
\newcommand{\laststep}{\global\PS@laststeptrue}
\newif\ifPS@lastsubstep
\newcommand{\lastsubstep}{\global\PS@lastsubsteptrue}

\newcommand{\adjustcol}[1]{%
  \global\advance\@colroom-#1%
}

\makeatother

\title{Forcing, Transition Algebras, and Calculi}
\author{Hashimoto Go}%
{Kyushu University, Japan}%
{hashimoto.go.427@s.kyushu-u.ac.jp}%
{}%
{}
\author{Daniel G\u{a}in\u{a}}%
{Kyushu University, Japan}%
{daniel@imi.kyushu-u.ac.jp}%
{}%
{}
\author{Ionu\c{t} \c{T}u\c{t}u}%
{Simion Stoilow Institute of Mathematics of the Romanian Academy, Romania}%
{ittutu@gmail.com}%
{}%
{}
\authorrunning{G.~Hashimoto, D.~G\u{a}in\u{a}, and I.~\c{T}u\c{t}u}
\Copyright{G.~Hashimoto, D.~G\u{a}in\u{a}, and I.~\c{T}u\c{t}u}
\funding{The work presented in this paper has been partially supported by Japan Society for the Promotion of Science, grant number 23K11048.}

\begin{document}

\keywords{Forcing, institution theory, calculi, algebraic specification, transition systems}

\maketitle

\begin{abstract}
  We bring forward a logical system of \emph{transition algebras} that enhances many-sorted first-order logic using features from dynamic logics.
  The sentences we consider include compositions, unions, and transitive closures of transition relations, which are treated similarly to the actions used in dynamic logics in order to define necessity and possibility operators.
  This leads to a higher degree of expressivity than that of many-sorted first-order logic.
  For example, one can finitely axiomatize both the finiteness and the reachability of models, neither of which are ordinarily possible in many-sorted first-order logic.
  We introduce syntactic entailment and study basic properties such as compactness and completeness, showing that the latter does not hold when standard finitary proof rules are used. 
  Consequently, we define proof rules having both finite and countably infinite premises, and we provide conditions under which completeness can be proved.
  To that end, we generalize the forcing method introduced in model theory by Robinson from a single signature to a category of signatures, and we apply it to obtain a completeness result for signatures that are at most countable.
\end{abstract}

\section{Introduction}
Algebraic specification is one of the main approaches to formal methods that supports both the formal specification of software and hardware systems and the formal verification of their requirements. 
The underlying logic of an algebraic specification language is often presented as an institution~\cite{gog-ins}, a category-theoretic formalization of the intuitive notion of logic that includes its syntax, semantics and the satisfaction relation between them. 
A lot of theoretical computer science has been developed within institution theory~\cite{Diaconescu2008,SannellaTarlecki2011,Diaconescu12} based on the principle that formal specification should be based rigorously upon a concrete institution.
Two notable specification languages have been designed by following this principle: CafeOBJ in Japan \cite{dia-caf} and CASL in Europe~\cite{DBLP:journals/tcs/AstesianoBKKMST02}.
However, there also is an exception given by the Maude system~\cite{DBLP:conf/maude/2007}, which was originally developed at SRI International in the United States. The underlying logic of Maude, called rewriting logic~\cite{Meseguer90}, is not given as an institution, which has led to a series of developments that diverged from mainstream institution-theoretic approaches to topics such as modularization and heterogeneity~\cite{CodescuMRM10}.
 
\minisec{Motivation}
The main goal of the present study is 
to apply a body of methods and principles developed within institutional model theory for 
defining a denotational semantics of algebraic specification languages that are executable by term rewriting such that:
\begin{enumerate}[nosep]
\item it enjoys the modular properties of the logic underlying CafeOBJ such as the \emph{satisfaction condition} for signature morphisms, and therefore it can be formalized as an \emph{institution};
\item it has the rich expressivity of rewriting logic, in the sense that it can provide a semantics for the Maude language, in general, and for its strategy language~\cite{EKER2023100887}, in particular.
\end{enumerate}
In algebraic specification languages executable by rewriting such as Maude and CafeOBJ, systems are specified using two kinds of atomic statements:
\begin{enumerate*}[label=(\textit{\alph*})]
\item \emph{equations}, which define an algebraic structure on system states, with constructors and derived operations, for example; and
\item \emph{transition rules}, which capture the behaviour of a system by telling us how the states may change as a result of certain actions.
\end{enumerate*}
In the present contribution, we propose a logic of \emph{transition algebras} where the models consist of many-sorted algebras equipped with binary relations that give semantics to the transition rules.
From transition rules one can construct \emph{actions} by applying composition, union, and the Kleene star (i.e., the reflexive and transitive closure of a relation).
For the sake of simplicity, we omit the subsorting relation~\cite{GM92}.

\minisec{Many-sorted logical systems}
Many-sorted logics are widely acknowledged as being suitable for applications in computer science.
However, in pure mathematical logic, they tend to be classified as ``inessential variation[s]'' \cite{monk76} of their unsorted forms. 
This might be true w.r.t.\ some classical aspects such as compactness or axiomatizability. However, in general, moving from the unsorted to the many-sorted case is a far from trivial task.
Allowing for multiple sorts, and thus for multiple carriers in models, some of which may be empty, alters the properties of the logics and significantly increases the complexity of proofs.

An important example of logical property that does not have a straightforward many-sorted generalization is Craig interpolation~\cite{Craig1957-CRALRA}.
This property generally holds in unsorted first-order logic, but fails to hold in the many-sorted variant of the logic; a counterexample can be found, for example, in \cite{DBLP:journals/ipl/Borzyszkowski00}.
Finding the most general criteria for Craig interpolation property was an open problem originally stated in \cite{tar-bit}.
A solution based on techniques advanced in institutional model theory was provided in \cite{gai-pop-rob} after nearly two decades.

Moreover, as noticed in \cite{gog-com}, if we admit models with potentially empty carrier sets, then proof rules for unsorted (or single-sorted) first-order logic may be unsound for its many-sorted counterpart. This already suggests that generalizations to other variants of many-sorted first-order logic may pose difficulties.
The completeness results proved in an institutional setting, such as \cite{Petria07,gai-comp,gai-acm} are applicable to logical systems where models interpret sorts as non-empty sets.
In fact, we are not aware of any completeness result for many-sorted first-order logic in which models interpret sorts as possibly empty sets.

\minisec{Forcing}
In the present contribution, we prove the completeness of the many-sorted logic of transition algebras by applying \emph{forcing}.
This technique was originally introduced by Paul Cohen~\cite{cohen63,cohen64} in set theory to show the independence of the continuum hypothesis from the other axioms of Zermelo-Fraenkel set theory. 
Robinson~\cite{rob71} developed an analogous theory of forcing in model theory.
In our setting, forcing is a technique used for constructing expanded models of consistent sets of sentences.
More specifically, it allows one to expand a set of sentences while preserving satisfiability even if compactness does not hold in the underlying logical system.
Transition algebra is not compact for the same reason classical dynamic propositional logic is not compact.
Therefore, the classical Henkin method for proving completeness, which relies on compactness, is not applicable to transition algebra.
Another issue arises when proving completeness from the fact that we work with models with empty carriers, because the addition of constants does not necessarily preserve the consistency of theories.
Therefore, we generalize the forcing method from a single signature to a category of signatures using ideas from institutional model theory such that the so-called Henkin constants can be added when needed in a way that preserves consistency.

\section{Transition algebra} \label{section:transition-algebra}
In this section, we define \emph{the logic of many-sorted transition algebras}, or \emph{transition algebra}~($\TA$), for short.
We present, in order: signatures, models, sentences, and the \(\TA\) satisfaction relation.
\minisec{Signatures}
The signatures we consider are ordinary algebraic signatures endowed with polymorphic transition labels and monotonic function symbols.
We denote them by tuples of the form \((S, F \supseteq M, L)\), where:
\begin{itemize}

\item \((S, F)\) is a many-sorted algebraic signature consisting of a set \(S\) of \emph{sorts} and a family \(F = \{ F_{w, s} \mid w \in S^{*}, s \in S \}\) of sets of \emph{function symbols};

\item \(M\) is a family of subsets \(M_{w, s} \subseteq F_{w, s}\) of \emph{monotonic} function symbols; and

\item \(L\) is a set whose elements we call \emph{transition labels}.

\end{itemize}

We often write \(\sigma \colon w \to s \in F\) to indicate that \(\sigma \in F_{w, s}\), and we refer to \(w \in S^{*}\) and \(s \in S\) as the \emph{arity} and \emph{sort}, respectively, of the symbol \(\sigma\).
Under this notation, \(F\) can also be regarded as an ordinary set consisting of \emph{function declarations} of the form \(\sigma \colon w \to s\).
When \(w\) is the empty arity, we may speak of \(\sigma \colon \to s\) as a \emph{constant} (symbol) of sort \(s\).

Throughout the paper, we let \(\Sigma\), \(\Sigma'\), and \(\Sigma_{i}\) range over arbitrary signatures of the form \((S, F \supseteq M, L)\), \((S', F' \supseteq M', L')\), and \((S_{i}, F_{i} \supseteq M_{i}, L_{i})\), respectively.

As usual in institution theory~\cite{Diaconescu2008,SannellaTarlecki2011}, important constructions such as signature extensions with constants as well as open formulae and quantifiers are realized in a multi-signature setting, so moving between signatures is common.
A \emph{signature morphism} \(\chi \colon \Sigma \to \Sigma'\) consists of
an ordinary algebraic signature morphism \(\chi \colon (S, F) \to (S', F')\)
such that \(\chi(M) \subseteq M'\) together with a function \(L \to L'\), which we typically denote using the same symbol, \(\chi\).

\begin{remark}
  Signature morphisms compose componentwise.
  Their composition has identities and is associative, thus leading to a category $\Sig$ of signatures.
\end{remark}

\minisec{Models}
Given a signature \(\Sigma\), a \emph{\(\Sigma\)-model} \(\mathfrak{A}\) is an \((S, F)\)-algebra \(\mathfrak{A}\) that interprets every label \(\lambda \in L\) as a \emph{many-sorted transition relation} $\lambda^\A \subseteq \A \times \A$ 
(that is,  $\lambda^\A=\{\lambda^\A_s\mid s\in S\}$ and $\lambda^\A_s\subseteq \A_s\times \A_s$ for all sorts $s\in S$)
that respects monotonic function symbols 
(that is, for all function symbols \(\sigma \colon s_{1} \dotsm s_{n} \to s\) in \(M\), all tuples \( (a_1,\dots,a_n)  \in \A_{s_1} \times \dots \times \A_{s_n}\), all indices \( k\in \{1,\dots,n\}\), and all elements \(b \in \A_{s_k}\), 
if $\pos{a_k,b} \in \lambda^\A_{s_{k}}$ then $\pos{\sigma^\A(a_1\dots,a_k,\dots,a_n),  \sigma^\A(a_1\dots,b,\dots,a_n)}\in \lambda^\A_{s}$).

A \emph{homomorphism} $h:\A\to \B$ over a signature $\Sigma$ is an algebraic \((S, F)\)-homomorphism that preserves transitions: $h(\lambda^\A)\subseteq \lambda^\B$ for all \(\lambda \in L\).
It is easy to see that $\Sigma$-homomorphisms form a category, which we denote by $\Mod(\Sigma)$, under their obvious componentwise composition.

\begin{remark}
  Every signature morphism \(\chi \colon \Sigma \to \Sigma\) determines a \emph{model-reduct functor} $\_\red_\chi\colon\Mod(\Sigma')\to\Mod(\Sigma)$ such that:
  \begin{itemize}

  \item for every \(\Sigma'\)-model \(\A'\), \((\A'\red_\chi)_s=\A'_{\chi(s)}\) for each sort \(s \in S\), \(\sigma^{\A'\red_{\chi}} = \chi(\sigma)^{\A'}\) for each symbol \(\sigma \in F\), and \(\lambda^{\A'\red_{\chi}} = \chi(\lambda)^{\A'}\) for each label \(\lambda \in L\); and

  \item for every \(\Sigma'\)-homomorphism \(h' \colon \A' \to \B'\), \((h' \red_{\chi})_{s} = h'_{\chi(s)}\) for each \(s \in S\).
    
  \end{itemize}
  Moreover, the mapping \(\chi \mapsto \_\red_\chi\) is functorial.
\end{remark}

For any signature morphism $\chi:\Sigma\to\Sigma'$, any $\Sigma$-model $\A$ and any $\Sigma'$-model $\A'$ if $\A=\A'\red_\chi$, we say that $\A$ is the \emph{$\chi$-reduct} of $\A'$, and that $\A'$ is a \emph{$\chi$-expansion} of  $\A$.
For example, for any many-sorted set \(X\) (say, of variables) that is disjoint from the set of constant-function symbols in \(\Sigma\), consider the inclusion morphism $\iota_X:\Sigma\hookrightarrow\Sigma[X]$, where \(\Sigma[X] = (S, F[X] \supseteq M, L)\) is the signature obtained from \(\Sigma = (S, F \supseteq M, L)\) by adding the elements of \(X\) to \(F\) as new constant-operation symbols of appropriate sort.
Then an expansion of a $\Sigma$-model $\A$ along $\iota_X$ can be seen as a pair $\pos{\A,g:X\to \A}$, where $g$ is a valuation of $X$ in $\A$.

As in many-sorted algebra, there is a special, initial model in \(\Mod(\Sigma)\), which we denote by \(T_{\Sigma}\), whose elements are ground terms built from function symbols, and whose transition relations are all empty.
The \(\Sigma\)-model \(T_{\Sigma}(X)\) of terms with variables from \(X\) is defined as the $\iota_X$-reduct of $T_{\Sigma[X]}$; i.e., $T_\Sigma(X)=T_{\Sigma[X]}\red_{\iota_X}$.
The following property is an immediate consequence of the initiality of \(T_{\Sigma}\).

\begin{remark}
  Any signature morphism \(\chi \colon \Sigma \to \Sigma'\) determines uniquely a \(\Sigma\)-ho\-mo\-mor\-phism \(T_{\Sigma} \to T_{\Sigma'} \red_{\chi}\).
  In order to simplify notations later on, we denote that homomorphism by \(\chi \colon T_{\Sigma} \to T_{\Sigma'} \red_{\chi}\);
  therefore, for any \(\Sigma\)-term \(\sigma(t_{1}, t_{2}, \dotsc, t_{n})\), we have \(\chi(\sigma(t_{1}, t_{2}, \dotsc, t_{n})) = \chi(\sigma)(\chi(t_{1}), \chi(t_{2}), \dotsc, \chi(t_{n}))\).
\end{remark}
\minisec{Sentences}
The \emph{actions} over a signature $\Sigma$ are defined by the following grammar:
\begin{center}
$\act\Coloneqq \lambda \mid 
\act\comp\act \mid 
\act\cup\act \mid 
\act^*$
\end{center}
where $\lambda$ is a transition label of \(\Sigma\).
We let \(A\) denote the set of all actions obtained from transition labels declared in a signature \(\Sigma\), and we extend the notational convention that we use for the components of signatures to their corresponding sets of actions; that is, we usually denote by \(A'\) the set of actions over a signature \(\Sigma'\), by \(A_{i}\) the set of actions over a signature \(\Sigma_{i}\), and so on.
Moreover, through a slight abuse of notation, we also denote by \(\chi \colon A \to A'\) the canonical map determined by a signature morphism \(\chi \colon \Sigma \to \Sigma'\).

To define sentences, we assume a countably infinite set of \emph{variable names} $\{v_i\mid i<\omega\}$.
A \emph{variable} for a signature $\Sigma$ is a triple $\pos{v_i,s,\Sigma}$, where $v_i$ is a variable name and $s$ is a sort in $\Sigma$ -- the third component is used only to ensure that variables are distinct from the constant-operation symbols declared in \(\Sigma\), which is essential when dealing with quantifiers.
The set $\Sen(\Sigma)$ of \emph{sentences} over $\Sigma$ is given by the following grammar:
\begin{center}
  $\phi \Coloneqq  t_{1} = t_{2} \mid t_{1} \stackrel{\act}\Rightarrow t_{2} \mid \neg\phi \mid \bigvee\Phi \mid \Exists{X}\phi'$
\end{center}
where
\begin{enumerate*}[label=(\alph*)]
\item~$t_{1}$ and $t_{2}$ are \((S, F)\)-terms of the same sort;
\item~$\act\in A$ is an action;
\item~$\Phi$ is a finite set of $\Sigma$-sentences; and
\item~$X$ is a finite set of variables for $\Sigma$ and $\phi'$ is a $\Sigma[X]$-sentence.
\end{enumerate*}

When \(\Phi = \{\phi_{1}, \phi_{2}, \dotsc, \phi_{n}\}\), we may write \(\phi_{1} \vee \phi_{2} \vee \dotsb \vee \phi_{n}\) instead of \(\bigvee \Phi\).
Besides the above core connectives, we also make use of the following convenient (and standard) abbreviations:
$\bigwedge\Phi\coloneqq\neg\bigvee\{\neg\phi\mid\phi\in\Phi\}$ for finite conjunctions;
$\bot\coloneqq\bigvee\emptyset$ for falsity;
$\top\coloneqq\bigwedge\emptyset=\neg\bot$ for truth;
$\phi_1\rightarrow \phi_2\coloneqq \neg\phi_1\vee\phi_2$ for implications; and
\(\Forall{X} \phi' \coloneqq \neg \Exists{X} \neg \phi'\) for universally quantified sentences.

\begin{remark}
Any signature morphism $\chi\colon\Sigma\to\Sigma'$ can be canonically extended to a \emph{sentence-translation function} $\chi\colon\Sen(\Sigma)\to\Sen(\Sigma')$ given by:
\begin{itemize}[itemsep=1ex]
\item $\chi(t_{1} = t_{2}) = (\chi(t_{1})=\chi(t_{2}))$; 
\item $\chi(t_{1} \stackrel{\act}\Rightarrow t_{2}) = \chi(t_{1}) \stackrel{\chi(\act)}\Longrightarrow \chi(t_{2})$;
\item $\chi(\neg\phi) = \neg\chi(\phi)$; 
\item $\chi(\bigvee\Phi) = \bigvee\chi(\Phi)$; and
\item $\chi(\Exists{X}\phi') = \Exists{X'}\chi'(\phi')$, where \(X' = \{ \pos{x,\chi(s),\Sigma'} \mid \pos{x,s,\Sigma} \in X \}\) and $\chi':\Sigma[X]\to\Sigma'[X']$ is the extension of $\chi$ mapping each variable $\pos{x,s,\Sigma}\in X$ to $\pos{x,\chi(s),\Sigma'}\in X'$.

\end{itemize}
Moreover, this sentence-translation mapping is functorial in \(\chi\).
\end{remark}

For the sake of simplicity, we identify variables only by their name and sort, provided that there is no danger of confusion.
Using this convention, each inclusion morphism $\iota\colon\Sigma\hookrightarrow\Sigma'$ determines an inclusion function $\iota\colon\Sen(\Sigma)\hookrightarrow\Sen(\Sigma')$, which corresponds to the approach of classical model theory.
This simplifies the presentation greatly. 
A situation when we cannot apply this convention arises when translating a $\Sigma$-sentence $\Exists{X}\phi$ along the inclusion $\iota_X:\Sigma\hookrightarrow\Sigma[X]$.

\minisec{Satisfaction relation}
Actions are interpreted as binary transition relations in models.
Given a model $\A$ over a signature $\Sigma$, and actions $\act,\act_1,\act_2\in A$, we have:
\begin{itemize}
\item $(\act_1\comp\act_2)^\A=\act_1^\A \comp \act_2^\A$ (i.e., diagrammatic composition of binary relations);
\item $(\act_1\cup\act_2)^\A=\act_1^\A \cup \act_2^\A$ (the union of binary relations); and
\item $(\act^*)^\A= (\act^\A)^*$ (the reflexive and transitive closure of binary relations).
\end{itemize}

We define the \emph{satisfaction relation} between models and sentences as follows:
\begin{itemize}
\item $\A\models t_1=t_2$ iff $t_1^\A=t_2^\A$;
\item $\A\models t_1\stackrel{\act}\Rightarrow t_2$ iff $(t_1^\A, t_2^\A) \in \act^{\A}$;
\item $\A\models \neg\phi$ iff $\A\not\models\phi$,
\item $\A\models\bigvee\Phi$ iff $\A\models\phi$ for some sentence $\phi\in \Phi$, and
\item $\A\models\Exists{X}\phi'$ iff $\A'\models\phi'$ for some expansion $\A'$ of $\A$ to the signature $\Sigma[X]$.
\end{itemize}

For the sake of simplicity, we write $d_1 \stackrel{\act}\Longrightarrow d_2$ if $\pos{d_1,d_2}\in \act^\A$.

Let $\phi,\phi'$ be sets of $\Sigma$-sentences, $\A$ a $\Sigma$-model.
We also use the following notations:
\begin{itemize}
\item $\A\models\Phi$ iff $\A\models\phi$ for all sentences $\phi\in\Phi$;
\item $\Gamma\models\Phi$ iff $\A\models\Gamma$ implies $\A\models\Phi$ for all $\Sigma$-models $\A$.
\end{itemize}
In particular, we write $\Gamma\models\phi$ instead of $\Gamma\models\{\phi\}$ for any set of sentences $\Gamma$ and any single sentence $\phi$.
\begin{proposition}\label{proposition:trans_models}
For all signature morphisms $\chi:\Sigma\to\Sigma'$,
all $\Sigma'$-models $\A$ and
all sentences $\phi\in\Sen(\Sigma)$
we have:
$\A\red_{\chi}\models\phi
  \quad\text{iff}\quad
  \A\models\chi(\phi)$.
\end{proposition}

\begin{example}[CCS]
  To illustrate the expressivity of transition algebra, we refer to Robin Milner's \emph{calculus of communicating systems} (CCS)~\cite{Milner80,Milner89}, which is emblematic of a broad family of formal languages used for modelling and reasoning about concurrency.
  In a nutshell, CCS is a process calculus that enables syntactic descriptions of concurrent systems to be written, and subsequently manipulated and analysed, based on two kinds of atomic entities -- process identifiers and channel names -- and a handful of composition operators.

  To start, we assume two sets: \(\mathit{PI}\) of \emph{process identifiers}, and \(\mathit{CN}\) of so-called \emph{channel names}, which capture the interaction capabilities of processes.
  Take, for instance, the following famous quote by Alfr\'{e}d R\'{e}nyi: ``A mathematician is a machine for turning coffee into theorems''~\cite{Schechter00}.
  This can be modelled in CCS as an interaction between two processes, a mathematician and a coffee vending machine, that trade coffee (in exchange, perhaps, of coins or some other form of payment) in order to jointly produce theorems.
  Hence, we can consider theorems, coffee, and coins as types of interactions between the two processes.

  For each channel name \(c \in \textit{CN}\), we let \(\overline{c}\) be a new symbol, distinct from all channel names, called the \emph{co-name} of \(c\).
  We also let \(\overline{\mathit{CN}} = \{ \overline{c} \mid c \in \mathit{CN}\}\) be the set of all co-names, and \(L = \mathit{CN} \cup \overline{\mathit{CN}}\) be the set of \emph{labels}.
  Intuitively, we may regard the symbols in \(\mathit{CN}\) as inputs of some process, and the symbols in \(\overline{\mathit{CN}}\) as outputs.
  Besides labels, we also consider an additional \emph{silent-action} symbol, denoted \(\tau\), that indicates an internal, unobservable behaviour of the system under consideration.
  Altogether, we refer to the symbols in \(A=L \cup \{\tau\}\) as CCS \emph{actions}.\footnote{Although they share the name `action', CCS actions are conceptually very different from the actions we have defined for transition algebra. To distinguish the two, we always prefix the former by CCS.}
  \emph{Processes} over \(\mathit{CN}\) and \(\mathit{PI}\) are defined according to the following grammar:
  \begin{grammar}
    P \ceqdef 0
    \alt \pi
    \alt a \mathbin{.} P
    \alt P + P
    \alt P \mathbin{\text{`\(\mid\)'}} P
    \alt P \setminus k
  \end{grammar}
  where
  \begin{inlinenum}
  \item \(0\) denotes a special terminal, inactive process;
  \item \(\pi \in \mathit{PI}\) is a process identifier;
  \item \(a \in A\) is a CCS action, which can be used to prefix a process \(P\) in order to form a new process, \(a \mathbin{.} P\), that intuitively performs \(a\) then continues as \(P\);
  \item `\(+\)' denotes the non-deterministic choice between two processes;
  \item `\(\mid\)' denotes the parallel composition of two processes; and
  \item \(k \in \mathit{CN}\) is a channel name, which can be used in expressions like \(P \setminus k\) to form a new, restricted process with the same interaction capabilities as \(P\) except for the labels \(k\) and \(\overline{k}\).
  \end{inlinenum}
  To simplify the presentation, we omit the relabelling operator because it plays no role in the examples we consider in this paper and it could be added with ease if needed.
  For a comprehensive account of CCS, see for example~\cite{Milner89}.

  A CCS \emph{context}, or \emph{program}, is a set of \emph{declarations} of the form \(\pi \Coloneqq P\), where \(\pi \in \mathit{PI}\) is a process identifier and \(P\) is a process conforming to the grammar introduced above, such that any two distinct declarations have distinct left-hand sides; in other words, we do not admit multiple declarations of the same process identifier within a given context.

  Using this syntax, the interaction between mathematicians and coffee vending machines announced in  Alfr\'{e}d R\'{e}nyi's quote can be formalized as a parallel composition of processes over $\textit{PI}=\{\texttt{Mathematician},\texttt{CoffeeVM}\}$ and $\textit{CN}=\{\texttt{coin},\texttt{coffee},\texttt{theorem}\}$, which we write as
  \(\texttt{Mathematician} \mid \texttt{CoffeeVM}\), where \texttt{Mathematician} and \(\texttt{CoffeeVM}\) are process identifiers `defined' recursively according to the following context:
  \begin{itemize}
  \item \(\texttt{Mathematician} \Coloneqq \overline{\texttt{coin}} \mathbin{.} \texttt{coffee} \mathbin{.} \overline{\texttt{theorem}} \mathbin{.} \texttt{Mathematician}\),
  \item \(\texttt{CoffeeVM} \Coloneqq \texttt{coin} \mathbin{.} \overline{\texttt{coffee}} \mathbin{.} \texttt{CoffeeVM}\).
  \end{itemize}
  Thanks to the expressivity of transition algebra, we can easily capture both the syntax and the operational semantics of CCS.
  For the syntax of processes, it suffices to consider a many-sorted $\TA$ signature with \(S = \{\texttt{Channel}, \texttt{Action}, \texttt{Process}\}\) and with \(F\) given by the following function symbols (which employ OBJ's~\cite{Goguen2000} and Maude's~\cite{DBLP:conf/maude/2007} mixfix notation):
  \begin{itemize}
  \item \(0 \colon  \to \texttt{Process}\),
  \item \(\pi \colon  \to \texttt{Process}\) \quad for each process identifier \(\pi \in \textit{PI}\),
  \item \(a \colon \rightarrow \texttt{Action}\) \quad for each CCS action \(a \in A\),
  \item \(\underline{~} \mathbin{.} \underline{~} \colon \texttt{Action}~\texttt{Process} \to \texttt{Process}\),
  \item \(\underline{~} + \underline{~}~, ~\underline{~} \mid \underline{~} \colon \texttt{Process}\; \texttt{Process} \to \texttt{Process}\),
  \item \(k \colon \rightarrow \texttt{Channel}\) \quad for each channel name \(k \in \mathit{CN}\),\footnote{To avoid subsorting, we overload channel names, which can be seen either as constants of sort \texttt{Channel} or as constants of sort \texttt{Action} depending on the context in which they are used.}
  \item \(\underline{~} \setminus \underline{~} \colon \texttt{Process}~\texttt{Channel} \to \texttt{Process}\).
  \end{itemize}
  The parallel-composition operator is the only monotonic function symbol of this example.
  For convenience, we also declare the parallel-composition operator and the non-deterministic choice as associative, commutative, and with identity \(0\).
  These properties can be presented -- as usual in algebraic specification -- using plain equations.
  To capture and reason about the behaviour of processes, we regard each CCS action as an atomic action (i.e., a transition label) in transition algebra -- and those are the only \(\TA\) labels that we consider here.

  The transitional semantics of a CCS program \(\textit{Pgm}\) is given by the following collection of axioms.
  The transition-algebra sentences below are all universally quantified over variables \(P\), \(P'\), \(Q\), \(Q'\) of sort \texttt{Process} and \(k\) of sort \texttt{Channel}; however, we drop the quantifiers in order to simplify the notation.
  We also use names for the axioms (at the beginning of each line) that are indicative of the transition rules defined in~\cite{Milner89}.
  \begin{itemize}
  \item\label{axioms:CCS}
    \makebox[3em][l]{(\emph{Act})} \(a \mathbin{.} P \stackrel{a}{\Rightarrow} P\) \quad for all \(a \in A\),
  \item \makebox[3em][l]{(\emph{Sum})} \(P \stackrel{a}{\Rightarrow} P' \rightarrow P + Q \stackrel{a}{\Rightarrow} P'\) \quad for all \(a \in A\),
  \item \makebox[3em][l]{(\emph{Com})} \(P \stackrel{c}{\Rightarrow} P' \land Q \stackrel{\overline{c}}{\Rightarrow} Q' \rightarrow P \mid Q \stackrel{\tau}{\Rightarrow} P' \mid Q'\) \quad for all \(c \in \textit{CN}\),
  \item \makebox[3em][l]{(\emph{Res})} \(P \stackrel{a}{\Rightarrow} P' \wedge a \neq k \wedge \overline{a} \neq k \rightarrow P \setminus k \stackrel{a}{\Rightarrow} P' \setminus k\) \quad for all \(a \in A\),\footnote{As usual in CCS, we extend the over-line notation employed for channel co-names to a bijection \(\overline{\cdot} \colon A \to A\) given by \(\overline{\overline{c}} = c\) for all channel names \(c\) and by \(\overline{\tau} = \tau\) for the silent action.}
  \item \makebox[3em][l]{(\emph{Con})} \(P \stackrel{a}{\Rightarrow} P' \rightarrow \pi \stackrel{a}{\Rightarrow} P'\) \quad for all \(\pi \Coloneqq P \in \textit{Pgm}\).
  \end{itemize}

  To simplify some of the notations used later on in the paper, for any process \(P\) and any non-empty and finite sequence \(K = (k_{i} \mid 1 \leq i \leq n)\) of channel names, we also write \(P \setminus K\) in place of \(P \setminus k_{1} \setminus \dotsb \setminus k_{n}\) and we consider the following derived form of the axiom \((\mathit{Res})\):
  \begin{itemize}
  \item \makebox[3em][l]{(\(\text{\emph{Res}}^{*}\))} \(P \stackrel{a}{\Rightarrow} P' \wedge \bigwedge \{a \neq k_{i} \wedge \overline{a} \neq k_{i} \mid 1 \leq i \leq n\} \rightarrow P \setminus K \stackrel{a}{\Rightarrow} P' \setminus K\) \quad for all \(a \in A\).
  \end{itemize}
  
  Similar encodings of CCS in languages that support transitions can be found in the rewriting-logic literature, notably in~\cite{Marti-OlietM96,VerdejoM02,DeganoGP02}.
  But the encodings presented therein rely on a notion of \emph{derivative} of a process instead of reasoning about plain processes, which is usually because labelled transitions cannot be used in the conditions of Horn clauses such as \emph{Sum} and \emph{Com}.
  That is, the axiomatization is done in terms of pairs \(\langle \alpha, P \rangle\), where \(P\) is a process and \(\alpha\) is a CCS action or a sequence of CCS actions leading to \(P\).
  In \(\TA\), this additional step can be avoided because the use of labelled transitions is unrestricted, which allows our axioms to be nearly to-the-letter transcriptions of Robin Milner's rules for CCS.
\end{example}

\section{Entailment relations} \label{sec:entailments}
In this section, we define the proof-theoretic properties necessary for proving our results such as entailment relation, soundness and completeness.
Before we proceed, let us recall an example from \cite{DBLP:journals/sigplan/GoguenM82}, which shows that classical rules of first-order deduction are not sound.
\begin{example}\label{ex:sound}
Let $\Sigma=(S,F)$ be an algebraic signature consisting of:
\begin{itemize}
\item two sorts, that is, $S=\{Elt,Bool\}$, and 
\item five function symbols 
$F=\{true:\to Bool, 
false:\to Bool, 
\mathsf{\sim\!\_}:Bool\to Bool, 
\mathsf{\_\&\_}: Bool~Bool\to Bool, 
\mathsf{\_+\_}: Bool~Bool\to Bool,
foo:Elt\to Bool\}$.
\end{itemize}
Let $\Gamma$ be a set of sentences over $\Sigma$ which consists of the following sentences:
\begin{itemize}
\item $\sim true = false$ and  $\sim false = true$, 
\item $\Forall{y} y ~\& \sim y= false$ and $\Forall{y} y ~\&~ y = y$,
\item $\Forall{y} y ~+ \sim y= true$ and $\Forall{y} y ~ + ~ y = y$, and
\item $\Forall{x}\sim foo(x)=foo(x)$.
\end{itemize}
Using the ordinary rules of first-order deduction, we can show that
\begin{equation} \label{eq:sound}
\begin{array}{l l}
true & = foo(x) + \sim foo(x)\\
& = foo(x) + foo(x)\\
& = foo(x) \\
& = foo(x) ~\&~ foo (x) \\
& = foo(x) ~\& \sim foo(x) \\
& = false 
\end{array}
\end{equation}
As a result, one would expect $true = false$ to hold in all algebras satisfying $\Gamma$. 
But that is not the case.
To see why, suppose $\A$ is the algebra obtained from $T_\Sigma$ through a factorization under the congruence relation $\equiv_\Gamma$ generated by $\Gamma$, that is:
\begin{itemize}
\item $\A_{Bool}=\{true/_{\equiv_\Gamma},false/_{\equiv_\Gamma}\}$ and $\A_{Elt}=\emptyset$,
\item $\sim$ is interpreted as the negation, $\&$ as the conjunction and $+$ as the disjunction, and
\item $foo^\A$ is the empty function.
\end{itemize}
Clearly, the algebra \(\A\) satisfies the sentences in \(\Gamma\) referring to the negation, conjunction, and disjunction of booleans.
Moreover, since there is no function from $\{x\}$ to $\A_{Elt}=\emptyset$, we have $\A \models_\Sigma \Forall{x}\sim foo(x)=foo(x)$.
It follows that $\A\models_\Sigma \Gamma$ but $\A\not\models_\Sigma true = false$.
\end{example}

This shows that moving from the unsorted to the many-sorted case is not as straightforward as one might expect.
Since a model may have some empty domains, one needs to design proof rules that take into account changes of signatures.

\begin{definition} [Entailment relation]
An entailment relation $\vdash=\{\vdash_\Sigma\}_{\Sigma\in|\Sig^\TA|}$ is a family of binary relations between sets of sentences indexed by signatures, that is, $\vdash_\Sigma \subseteq \mathcal{P}(\Sen(\Sigma))\times \mathcal{P}(\Sen(\Sigma))$ for all first-order signatures $\Sigma$, such that the following properties are satisfied: \smallskip

\noindent
\begin{tabular}{l l l}
$(Monotonicity)~\proofrule{\Gamma\supseteq \Phi
}{ \Gamma\vdash_\Sigma \Phi}$ & &
$(Transitivity)~\proofrule{\Gamma\vdash_\Sigma \Phi \Space \Phi\vdash_\Sigma \Psi}{\Gamma\vdash_\Sigma \Psi}$ \\
&&\\
$(Union)~\proofrule{\Gamma\vdash_\Sigma \phi \text{ for all } \phi\in\Phi}{\Gamma\vdash_\Sigma \Phi}$ && 
$(Translation)~\proofrule{\Gamma\vdash_{\Sigma}\Phi}{\chi(\Gamma)\vdash_{\Sigma'}\chi(\Phi)}$ where $\chi:\Sigma\to\Sigma'$\\
\end{tabular}
\end{definition}
For the sake of simplicity, we write $\Gamma\vdash_\Sigma\phi$ rather than $\Gamma\vdash_{\Sigma}\{\phi\}$.
Also, we omit the subscript $\Sigma$ from the notation $\vdash_\Sigma$ when it is clear from the context.
An example of entailment relation is $\models$.
It is straightforward to prove that $\models$ satisfies $(Monotonicity)$, $(Transitivity)$, $(Union)$ and $(Translation)$.
\begin{definition}[Entailment properties]
An entailment relation $\vdash$ is \emph{sound} $($\emph{complete}$)$ if $\vdash\ \subseteq\ \models$ $(\models\ \subseteq\ \vdash)$.
An entailment relation $\vdash$ is $\alpha$-\emph{compact}, where $\alpha$ is an infinite cardinal, if
\begin{center}
$\Gamma\vdash_\Sigma\phi$ implies $\Gamma_\alpha\vdash_\Sigma\phi$ for some subset $\Gamma_\alpha\subseteq \Gamma$ of cardinality $\card(\Gamma_\alpha)< \alpha$,
\end{center}
for all signatures $\Sigma$, all sets of $\Sigma$-sentences $\Gamma$ and all $\Sigma$-sentences $\phi$.
If $\alpha=\omega$, we say, simply, that $\vdash$ is compact. 
\end{definition}
The dynamic entailment relation is defined in two steps.
Firstly, we define an entailment relation  to reason about the logical consequences of atomic sentences, given as equations or relations.
Secondly, we define the dynamic entailment relation by adding proof rules to deal with actions, Boolean connectives and quantifiers.

\subsection{Basic entailment relation}
The fragment obtained from $\TA$ by restricting the sentences to atoms is studied in this section.

\begin{definition}[Basic entailment relation] \label{def:basic}
\emph{The basic entailment relation} $\vdash^b$ is the least entailment relation closed under the following basic proof rules:
\end{definition}

\noindent
\setlength{\tabcolsep}{.5em}
\begin{tabular}{l l l}
$(R)~\proofrule{}{\Gamma\vdash_\Sigma t=t }$ 
&
$(S)~\proofrule{\Gamma\vdash_\Sigma t_1 = t_2}{\Gamma \vdash_\Sigma  t_2=t_1}~~ $ 
& 
$(T)~\proofrule{\Gamma\vdash_\Sigma t_1 = t_2 \Space \Gamma\vdash_\Sigma t_2 = t_3}{\Gamma \vdash_\Sigma  t_1 = t_3}$\\ 
&& \\
$(F)~\proofrule{\Gamma\vdash_\Sigma t_{i}=t_{i}' \quad \text{for}\ 1 \leq i \leq n}{\Gamma \vdash_\Sigma \sigma(t_{1}, \dotsc, t_{n})=\sigma(t_{1}', \dotsc, t_{n}')}$ ~~ & 
\multicolumn{2}{l}{
$(P)~\proofrule{\Gamma\vdash_\Sigma t_{1}=t'_{1} ~~~ \Gamma\vdash_\Sigma t_{2}=t'_{2} ~~~ \Gamma\vdash_\Sigma t_1\stackrel{\lambda}\Longrightarrow t_2} 
{\Gamma \vdash_\Sigma t'_{1}\stackrel{\lambda}\Longrightarrow t'_{2}}$
} \\
&& \\
\multicolumn{3}{l}{
$(M)~\proofrule{\Gamma\vdash_\Sigma t_{j}\stackrel{\lambda}\Longrightarrow u_{j}}{\Gamma \vdash_\Sigma  
f(t_{1},\dots,t_{j},\dots,t_{n})\stackrel{\lambda}\Longrightarrow f(t_{1},\dots,u_{j},\dots,t_{n})}$ ~~ where $f\in M$}
\end{tabular}
\begin{lemma}[Basic compactness] \label{lemma:basic-compact}
The basic entailment relation is compact.
\end{lemma}
Any set of atomic sentences $E$ defined over a signature $\Sigma$ determines a congruence 
$\equiv^E\coloneqq\{t_1 \equiv^E t_2\mid  E \vdash_\Sigma t_1 = t_2\}$ on $T_\Sigma$.
One can construct a model $\A^E$ from the initial model of terms $T_\Sigma$ factorized by the congruence $\equiv^E$ interpreting each transition label $\lambda$ in $\Sigma$ as the set $\{(t_1,t_2)\mid t_1,t_2\in T_{\Sigma,s} \text{ and } E\vdash_\Sigma t_1 \stackrel{\lambda}{\Longrightarrow} t_2\}$.
\begin{lemma}\label{lemma:basic}
Let $E$ be a set of atomic sentences defined over a signature $\Sigma$.
For all $\Sigma$-models $\A$, we have $\A\models E$ iff there exists a unique homomorphism $\A^E\to \A$.
\end{lemma}
Lemma~\ref{lemma:basic} says that the satisfaction of $E$ by a model $\A$ is equivalent with the existence of a unique homomorphism from $\A^E$ to $\A$.
In particular, $\A^E$ is the initial model of $E$.
See \cite{Diaconescu2008} for a proof of Lemma~\ref{lemma:basic}.
\begin{proposition} [Basic completeness] \label{prop:basic-complete} 
For any set of atomic sentences $E$ and any atomic sentence $\varphi$ defined over a signature $\Sigma$, the following are equivalent: \smallskip

\centering
\begin{enumerate*}[label=(\alph*)]
\item $E\models \varphi$, ~~~
\item $\A^E\models\varphi$, ~~~ and ~~~
\item $E\vdash^b \varphi$.
\end{enumerate*}
\end{proposition}
\subsection{Dynamic entailment relation}
The dynamic entailment relation is built on top of basic entailment relation by adding the proof rules to reason about
actions,
Boolean connectives, and
first-order quantifiers.
\begin{definition}[Dynamic entailment relation] \label{def:dynamic-rules}
\emph{The dynamic entailment relation} $\vdash$ is the least entailment relation closed under the basic proof rules presented in Definition~\ref{def:basic} and the following proof rules:
\end{definition}

\noindent 
\textbf{Proof rules for actions}\\
\begin{tabular}{l}
\\
$(Comp_I)~\proofrule{\Gamma\vdash_\Sigma t_1\stackrel{\act_1}\Longrightarrow  t \Space \Gamma\vdash_\Sigma t\stackrel{\act_2}\Longrightarrow t_2}{\Gamma\vdash_\Sigma t_1\stackrel{\act_1\comp\act_2}\Longrightarrow t_2}$ 
\\
\\		
$(Comp_E)~\proofrule{\Gamma\vdash_\Sigma t_1\stackrel{\act_1\comp\act_2}\Longrightarrow  t_2 ~~~ \Gamma \cup \{t_1\stackrel{\act_1}\Longrightarrow x,x\stackrel{\act_2}\Longrightarrow t_2\}\vdash_{\Sigma[x]}\phi}{\Gamma\vdash_\Sigma\phi}$ 
\\
\end{tabular}
\medskip

\noindent
\begin{tabular}{l l}
$(Union_I)~\proofrule{\Gamma \vdash_\Sigma t_1\stackrel{\act_i}\Longrightarrow t_2}{\Gamma\vdash_\Sigma t_1\stackrel{\act_1\cup\act_2}\Longrightarrow t_2}$ &
$(Union_E)~\proofrule{\Gamma\vdash_\Sigma t_1\stackrel{\act_1\cup\act_2}\Longrightarrow t_2 ~~ \Gamma\cup\{t_1\stackrel{\act_i}\Longrightarrow t_2\}\vdash_\Sigma \phi\text{ for all } i \in \{ 1,2 \}}{\Gamma\vdash_\Sigma \phi}$
\\
\\
$(Star_I)~\proofrule{\Gamma\vdash_\Sigma t_1\stackrel{\act^n}\Longrightarrow t_2}{\Gamma\vdash_\Sigma t_1\stackrel{\act^*}\Longrightarrow t_2}$ &
$(Star_E)~\proofrule{\Gamma\vdash_\Sigma t_1\stackrel{\act^*}\Longrightarrow t_2 ~~~ \Gamma\cup\{t_1\stackrel{\act^n}\Longrightarrow t_2\}\vdash_\Sigma \phi\text{ for all } n \in \omega}{\Gamma\vdash_\Sigma \phi}$
\end{tabular} \medskip

\noindent \textbf{Proof rules for Boolean connectives}\\
\begin{tabular}{l l}
& \\
$(Neg_D)~\proofrule{\Gamma\vdash_\Sigma \neg\neg \phi}{\Gamma\vdash_\Sigma\phi}$ &
$(False)~\proofrule{\Gamma\vdash_\Sigma \bot}{\Gamma\vdash_\Sigma\phi}$ \\
& \\
$(Neg_I)~\proofrule{\Gamma\cup\{\phi \}\vdash_\Sigma \bot}{\Gamma\vdash_\Sigma \neg\phi}$ &
$(Neg_E)~\proofrule{\Gamma\vdash_\Sigma\neg \phi}{\Gamma\cup\{\phi\} \vdash_\Sigma \bot}$\\
&\\
$(Disj_I)~\proofrule{\Gamma\vdash_\Sigma\phi}{\Gamma\vdash_\Sigma \vee\Phi}$ where $\phi\in \Phi$ &
$(Disj_E)~\proofrule{\Gamma\vdash_\Sigma\vee\Phi ~~~  \Gamma\cup\{\phi\}\vdash_\Sigma \gamma\text{ for all } \phi\in\Phi}{\Gamma\vdash_\Sigma \gamma}$
\end{tabular} \medskip

\noindent
\textbf{Proof rules for first-order quantifiers}\\
\begin{tabular}{l l}
&\\
$(Quant_I)~\proofrule{\Gamma \cup\{\phi\}\vdash_{\Sigma[X]}\gamma}{\Gamma\cup\{\Exists{X}\phi\}\vdash_\Sigma \gamma}$ ~~ &
$(Quant_E)~\proofrule{\Gamma\cup\{\Exists{X}\phi\}\vdash_\Sigma \gamma}{\Gamma\cup\{\phi\}\vdash_{\Sigma[X]}\gamma}$ 
  \\
& \\ 
$(Subst)~\proofrule{\Gamma \vdash_\Sigma \theta(\phi)}{\Gamma\vdash_\Sigma \Exists{X}\phi}$  & 
where $\theta:X\to T_{\Sigma}$ is a substitution\\
\end{tabular}

\begin{proposition} [$\omega_1$-compactness] \label{prop:omega-compact}
We have that
\begin{enumerate}
\item the dynamic entailment relation $\vdash$ is $\omega_1$-compact, and
\item the satisfaction relation $\models$ is not $\omega_1$-compact.
\end{enumerate}
\end{proposition}
The first statement holds because the dynamic entailment relation is generated by proof rules with an at most countable number of premises.
For uncountable signatures $\Sigma$, the satisfaction relation $\models_\Sigma$ is not $\omega_1$-compact.
It follows that the dynamic logic proposed in this contribution, $\TA$, is not complete. 
However, the restriction of $\TA$ to countable signatures, $\TA^c$, is complete.
Since $\TA^c$ is not compact, the Henkin method for proving completeness is not applicable. 

\begin{example}[Analysis of CCS programs]
  Recall the CCS description of the interaction between mathematicians and coffee vending machines discussed in Section~\ref{section:transition-algebra}, and let \texttt{Institute} be an abbreviation for the following process:
  \setlength{\abovedisplayskip}{1ex}
  \setlength{\belowdisplayskip}{1ex}
  \[
    (\texttt{Mathematician} \mid \texttt{CoffeeVM}) \setminus (\texttt{coin}, \texttt{coffee}).
  \]

  In this context, can we check, as an example, that the process \texttt{Institute} is able to continuously output theorems?
  The property can be formalized in \(\TA\) as a transition
  \[
    \texttt{Institute} \xRightarrow{\tau^{*}\, \comp \,\overline{\texttt{theorem}}\, \comp \,\tau^{*}} \texttt{Institute}
  \]
  whose \(\tau\)-components correspond to internal communications between sub-processes of the institute -- i.e., mathematicians and vending machines.
  Therefore, we need to check an entailment of the form \(\Gamma \vdash_{\Sigma} \phi\), where
  \begin{inlinenum}
  \item \(\Sigma\) is the \(\TA\)-signature that consists of the process identifiers \texttt{Mathematician} and \texttt{CoffeeVM} together with the CCS process-building operators for action prefixing, non-deterministic choice, parallel composition, etc., discussed in Section~\ref{section:transition-algebra};
  \item \(\Gamma\) is the set of \(\Sigma\)-sentences given by the axiom schemas \emph{Act}, \emph{Sum}, \emph{Com}, \(\text{\emph{Res}}^{*}\), and \emph{Con} listed on page~\pageref{axioms:CCS} together with equations pertaining to the axiomatization of CCS actions (e.g., \(\overline{\texttt{theorem}} \neq \texttt{coffee}\)), as well as equations that capture elementary properties of processes such as the associativity, commutativity, and identity element of the non-deterministic-choice and parallel-composition operators; and
  \item \(\phi\) is the transition written above.
  \end{inlinenum}

  The proof mimics the following chain of CCS transitions:
  \begin{flalign*}
    \texttt{Institute} & \xRightarrow{\tau} (\texttt{coffee} \mathbin{.} \overline{\texttt{theorem}} \mathbin{.} \texttt{Mathematician} \mid \overline{\texttt{coffee}} \mathbin{.} \texttt{CoffeeVM}) \setminus (\texttt{coin}, \texttt{coffee}) & \\[-5pt]
     & \xRightarrow{\tau} (\overline{\texttt{theorem}} \mathbin{.} \texttt{Mathematician} \mid \texttt{CoffeeVM}) \setminus (\texttt{coin}, \texttt{coffee}) & \\[-5pt]
     & \xRightarrow{\overline{\texttt{theorem}}} (\texttt{Mathematician} \mid \texttt{CofeeVM}) \setminus (\texttt{coin}, \texttt{coffee}) = \texttt{Institute} &
  \end{flalign*}
  To shorten the presentation of the proof, we use the following derived proof rule:
  \par\medskip\noindent
  \begin{tabular}{l l}
    $(\textit{GMP})~\proofrule{\Gamma\vdash_\Sigma \Forall{X}\bigwedge\Phi \rightarrow \gamma ~~~  \Gamma\vdash_\Sigma \theta(\Phi)}{\Gamma\vdash_\Sigma \theta(\gamma)}$ & where $\theta:X\to T_{\Sigma}$ is a substitution. \\
  \end{tabular}
  \par\medskip
  \noindent In addition, we simplify the notations by writing down only the conclusions of entailments and by abbreviating \texttt{Mathematician} as \texttt{M}, \texttt{CoffeeVM} as \texttt{CVM}, and the sequence of channel names \((\texttt{coin}, \texttt{coffee})\) as \(K\).
  This leads us to the (sketch of) proof tree depicted in Figure~\ref{proof-tree:theorems}.

  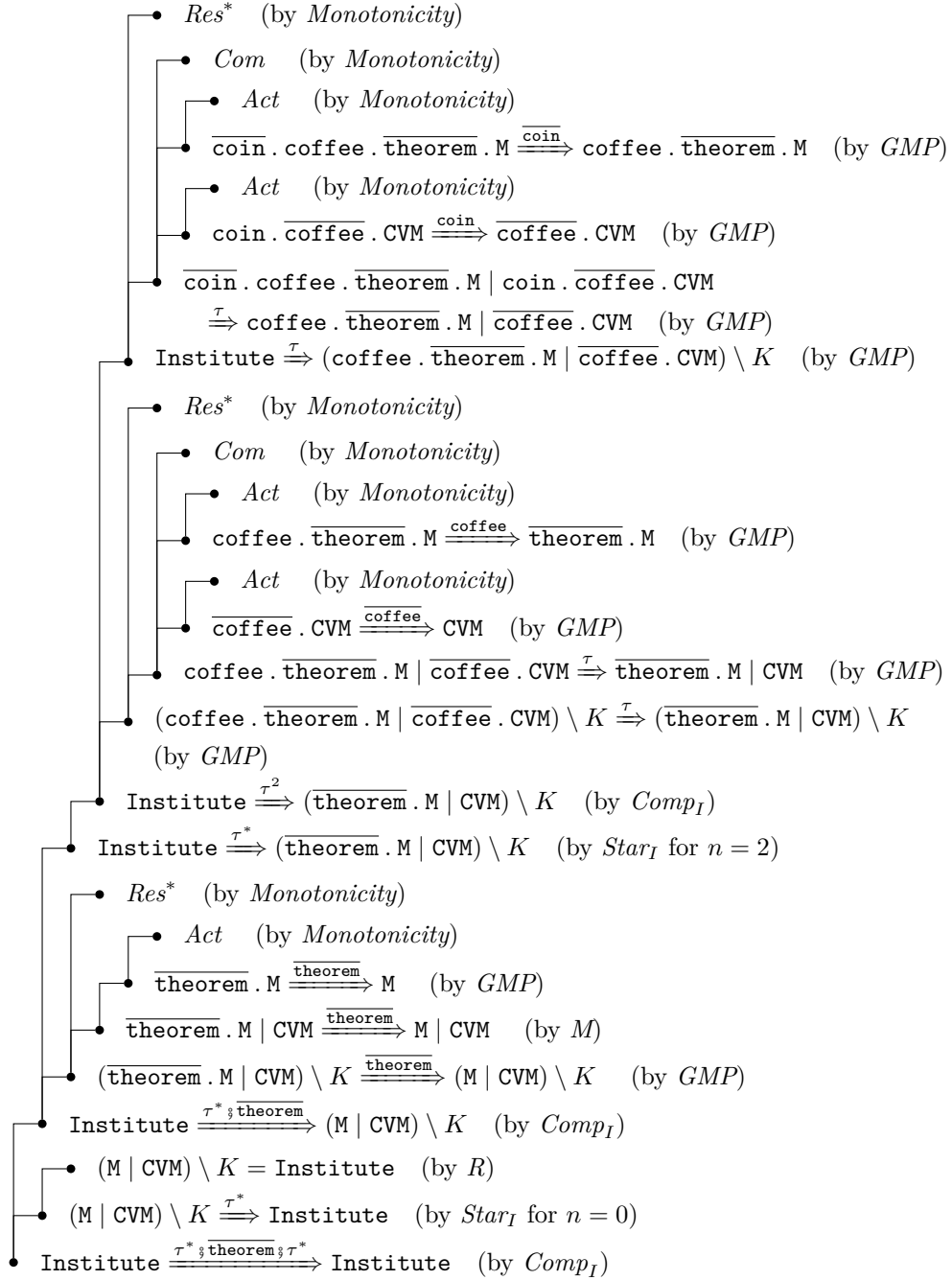
\begin{figure}
    \centering
    \begin{tikzpicture}[bullet/.style={fill, circle, anchor=center, inner sep=1.2pt, outer sep=0pt}, rule/.style={anchor=west, inner sep=1.5pt, outer sep=0pt}]
      \node [bullet] (E) {};
      \node [right=.5em of E] {\(\texttt{Institute} \xRightarrow{\tau^{*}\, \comp \,\overline{\texttt{theorem}}\, \comp \,\tau^{*}} \texttt{Institute}\)\quad (by \(\textit{Comp}_{I}\))};
      \node [bullet, above right=4ex and 1em of E.center] (E1) {};
      \draw (E) |- (E1);
      \node [right=.5em of E1] {\((\texttt{M} \mid \texttt{CVM}) \setminus K \xRightarrow{\tau^{*}}  \texttt{Institute}\)\quad (by \(\textit{Star}_{I}\) for \(n = 0\))};
      \node [bullet, above right=4ex and 1em of E1.center] (E11) {};
      \draw (E1) |- (E11);
      \node [right=.5em of E11] {\((\texttt{M} \mid \texttt{CVM}) \setminus K =  \texttt{Institute}\)\quad (by \textit{R})};
      \node [bullet, above=8ex of E1.center] (E2) {};
      \draw (E) |- (E2);
      \node [right=.5em of E2] {\(\texttt{Institute} \xRightarrow{\tau^{*}\, \comp \,\overline{\texttt{theorem}}} (\texttt{M} \mid \texttt{CVM}) \setminus K\)\quad (by \(\textit{Comp}_{I}\))};
      \node [bullet, above right=4ex and 1em of E2.center] (E21) {};
      \draw (E2) |- (E21);
      \node [right=.5em of E21] {\((\overline{\texttt{theorem}} \mathbin{.} \texttt{M} \mid \texttt{CVM}) \setminus K \xRightarrow{\overline{\texttt{theorem}}} (\texttt{M} \mid \texttt{CVM}) \setminus K\) \quad (by \(\textit{GMP}\))};
      \node [bullet, above right=4ex and 1em of E21.center] (E211) {};
      \draw (E21) |- (E211);
      \node [right=.5em of E211] {\(\overline{\texttt{theorem}} \mathbin{.} \texttt{M} \mid \texttt{CVM} \xRightarrow{\overline{\texttt{theorem}}} \texttt{M} \mid \texttt{CVM}\) \quad (by \(\textit{M}\))};
      \node [bullet, above right=4ex and 1em of E211.center] (E2111) {};
      \draw (E211) |- (E2111);
      \node [right=.5em of E2111] {\(\overline{\texttt{theorem}} \mathbin{.} \texttt{M} \xRightarrow{\overline{\texttt{theorem}}} \texttt{M}\) \quad (by \(\textit{GMP}\))};
      \node [bullet, above right=4ex and 1em of E2111.center] (E21111) {};
      \draw (E2111) |- (E21111);
      \node [right=.5em of E21111] {\(\textit{Act}\) \quad (by \(\textit{Monotonicity}\))};
      \node [bullet, above=12ex of E211.center] (E212) {};
      \draw (E21) |- (E212);
      \node [right=.5em of E212] {\(\textit{Res}^{*}\)\quad (by \(\textit{Monotonicity}\))};
      \node [bullet, above=20.5ex of E21.center] (E22) {};
      \draw (E2) |- (E22);
      \node [right=.5em of E22] {\(\texttt{Institute} \xRightarrow{\tau^{*}} (\overline{\texttt{theorem}} \mathbin{.} \texttt{M} \mid \texttt{CVM}) \setminus K\)\quad (by \(\textit{Star}_{I}\) for \(n = 2\))};
      \node [bullet, above right=4ex and 1em of E22.center] (E221) {};
      \draw (E22) |- (E221);
      \node [right=.5em of E221] {\(\texttt{Institute} \xRightarrow{\tau^{2}} (\overline{\texttt{theorem}} \mathbin{.} \texttt{M} \mid \texttt{CVM}) \setminus K\)\quad (by \(\textit{Comp}_{I}\))};
      \node [bullet, above right=7ex and 1em of E221.center] (E2211) {};
      \draw (E221) |- (E2211);
      \node [right=.5em of E2211] {\((\texttt{coffee} \mathbin{.} \overline{\texttt{theorem}} \mathbin{.} \texttt{M} \mid \overline{\texttt{coffee}} \mathbin{.} \texttt{CVM}) \setminus K \xRightarrow{\tau} (\overline{\texttt{theorem}} \mathbin{.} \texttt{M} \mid \texttt{CVM}) \setminus K\)};
      \node [below right=1ex and .5em of E2211] {(by \(\textit{GMP}\))};
      \node [bullet, above right=4ex and 1em of E2211.center] (E22111) {};
      \draw (E2211) |- (E22111);
      \node [right=.5em of E22111] {\(\texttt{coffee} \mathbin{.} \overline{\texttt{theorem}} \mathbin{.} \texttt{M} \mid \overline{\texttt{coffee}} \mathbin{.} \texttt{CVM} \xRightarrow{\tau} \overline{\texttt{theorem}} \mathbin{.} \texttt{M} \mid \texttt{CVM}\)\quad (by \(\textit{GMP}\))};
      \node [bullet, above right=4ex and 1em of E22111.center] (E221111) {};
      \draw (E22111) |- (E221111);
      \node [right=.5em of E221111] {\(\overline{\texttt{coffee}} \mathbin{.} \texttt{CVM} \xRightarrow{\overline{\texttt{coffee}}} \texttt{CVM}\)\quad (by \(\textit{GMP}\))};
      \node [bullet, above right=4ex and 1em of E221111.center] (E2211111) {};
      \draw (E221111) |- (E2211111);
      \node [right=.5em of E2211111] {\(\textit{Act}\) \quad (by \(\textit{Monotonicity}\))};
      \node [bullet, above right=12ex and 1em of E22111.center] (E221112) {};
      \draw (E22111) |- (E221112);
      \node [right=.5em of E221112] {\(\texttt{coffee} \mathbin{.} \overline{\texttt{theorem}} \mathbin{.} \texttt{M} \xRightarrow{\texttt{coffee}} \overline{\texttt{theorem}} \mathbin{.} \texttt{M}\)\quad (by \(\textit{GMP}\))};
      \node [bullet, above right=4ex and 1em of E221112.center] (E2211121) {};
      \draw (E221112) |- (E2211121);
      \node [right=.5em of E2211121] {\(\textit{Act}\) \quad (by \(\textit{Monotonicity}\))};
      \node [bullet, above right=20ex and 1em of E22111.center] (E221113) {};
      \draw (E22111) |- (E221113);
      \node [right=.5em of E221113] {\(\textit{Com}\) \quad (by \(\textit{Monotonicity}\))};
      \node [bullet, above=24ex of E22111.center] (E22112) {};
      \draw (E2211) |- (E22112);
      \node [right=.5em of E22112] {\(\textit{Res}^{*}\)\quad (by \(\textit{Monotonicity}\))};
      \node [bullet, above=32.5ex of E2211.center] (E2212) {};
      \draw (E221) |- (E2212);
      \node [right=.5em of E2212] {\(\texttt{Institute} \xRightarrow{\tau} (\texttt{coffee} \mathbin{.} \overline{\texttt{theorem}} \mathbin{.} \texttt{M} \mid \overline{\texttt{coffee}} \mathbin{.} \texttt{CVM}) \setminus K\)\quad (by \(\textit{GMP}\))};
      \node [bullet, above right=7ex and 1em of E2212.center] (E22121) {};
      \draw (E2212) |- (E22121);
      \node [right=.5em of E22121] {\(\overline{\texttt{coin}} \mathbin{.} \texttt{coffee} \mathbin{.} \overline{\texttt{theorem}} \mathbin{.} \texttt{M} \mid \texttt{coin} \mathbin{.} \overline{\texttt{coffee}} \mathbin{.} \texttt{CVM}\)};
      \node [below right=1ex and .5em of E22121] {\(\quad \xRightarrow{\tau} \texttt{coffee} \mathbin{.} \overline{\texttt{theorem}} \mathbin{.} \texttt{M} \mid \overline{\texttt{coffee}} \mathbin{.} \texttt{CVM}\)\quad (by \(\textit{GMP}\))};
      \node [bullet, above right=4ex and 1em of E22121.center] (E221211) {};
      \draw (E22121) |- (E221211);
      \node [right=.5em of E221211] {\(\texttt{coin} \mathbin{.} \overline{\texttt{coffee}} \mathbin{.} \texttt{CVM} \xRightarrow{\texttt{coin}} \overline{\texttt{coffee}} \mathbin{.} \texttt{CVM}\)\quad (by \(\textit{GMP}\))};
      \node [bullet, above right=4ex and 1em of E221211.center] (E2212111) {};
      \draw (E221211) |- (E2212111);
      \node [right=.5em of E2212111] {\(\textit{Act}\) \quad (by \(\textit{Monotonicity}\))};
      \node [bullet, above right=12ex and 1em of E22121.center] (E221212) {};
      \draw (E22121) |- (E221212);
      \node [right=.5em of E221212] {\(\overline{\texttt{coin}} \mathbin{.} \texttt{coffee} \mathbin{.} \overline{\texttt{theorem}} \mathbin{.} \texttt{M} \xRightarrow{\overline{\texttt{coin}}} \texttt{coffee} \mathbin{.} \overline{\texttt{theorem}} \mathbin{.} \texttt{M}\)\quad (by \(\textit{GMP}\))};
      \node [bullet, above right=4ex and 1em of E221212.center] (E2212121) {};
      \draw (E221212) |- (E2212121);
      \node [right=.5em of E2212121] {\(\textit{Act}\) \quad (by \(\textit{Monotonicity}\))};
      \node [bullet, above right=20ex and 1em of E22121.center] (E221213) {};
      \draw (E22121) |- (E221213);
      \node [right=.5em of E221213] {\(\textit{Com}\) \quad (by \(\textit{Monotonicity}\))};
      \node [bullet, above=24ex of E22121.center] (E22122) {};
      \draw (E2212) |- (E22122);
      \node [right=.5em of E22122] {\(\textit{Res}^{*}\)\quad (by \(\textit{Monotonicity}\))};
    \end{tikzpicture}
    \caption{Proof tree for the continuous output of theorems by the process \texttt{Institute}.}
    \label{proof-tree:theorems}
  \end{figure}
\end{example}
\section{Forcing}\label{sec:forcing}
In this section, we develop a forcing technique for proving completeness which extends in a non-straightforward way the classical forcing from one signature to a category of signatures.

\begin{definition}[Forcing property] \label{def:forcing}
\emph{A forcing property} is a tuple $\mathbb P=(P,\leq,\Delta,f)$, where:

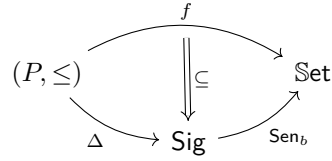
\centering
\begin{tikzcd}[row sep=small]
(P,\leq) \ar[dr,bend right=20,"\Delta",swap] \ar[rr,bend left,"f", ""{below,name=U}] &  & \Set\\
& \Sig \ar[ur,bend right=20,"\Sen_b",swap]  \ar[Rightarrow,from = U,"\subseteq"] & 
\end{tikzcd}
\flushleft

\begin{enumerate}
\item $(P,\leq)$ is a partially ordered set with a least element $0$.
		
The elements of $P$ are traditionally called conditions.
		
\item $\Delta: (P,\leq) \to \Sig$ is a functor, which maps each arrow $(p\leq q)\in (P,\leq)$ to an inclusion $\Delta_p\subseteq \Delta_q$.
\item $f:(P,\leq)\to\Set$ is a functor from the small category $(P,\leq)$ to the category of sets $\Set$  such that $f \subseteq \Delta;\Sen_b$ is a natural transformation, that is:
\begin{enumerate*}
\item $f(p)\subseteq \Sen_b(\Delta_p)$ for all conditions $p\in P$, and
\item $f(p)\subseteq f(q)$ for all arrows $(p\leq q)\in (P,\leq)$.
\end{enumerate*}
\item If $f(p)\models \phi$ then $\phi\in f(q)$ for some $q\geq p$, for all atoms $\phi\in \Sen_b(\Delta_p)$.
\end{enumerate}
\end{definition}
A classical forcing property is a particular case of forcing property such that $\Delta_p=\Delta_q$ for all conditions $p,q\in P$.
\begin{example}[Syntactic forcing] \label{ex:syntactic-forcing}
Let $\Sigma$ be a base signature and $C$ an $S$-sorted set of new constants such that $\card(C_s)=\omega$ for all $s\in S$.
Let $\mathbb{P}=(P,\leq,\Delta,f)$ be a forcing property defined as follows:
\begin{itemize}
\item $P$ is the set of presentations of the form $p=(\Delta_p,\Gamma_p)$,
where 
\begin{enumerate*}[label=(\alph*)]
\item $\Delta_p$ is obtained from $\Sigma$ by adding a finite set $C_p$ of constants from $C$, and  
\item $\Gamma_p\subseteq\Sen(\Delta_p)$ is consistent, that is, $\Gamma_p\not\vdash_{\Delta_p}\bot$.
\end{enumerate*}
\item $p\leq q$ iff $\Delta_p\subseteq\Delta_q$ and $\Gamma_p\subseteq \Gamma_q$, for all conditions $p,q\in P$.
\item $\Delta$ is the forgetful functor which maps each condition $p\in P$ to $\Delta_p$.
\item  $f(p)=\Gamma_p \cap \Sen_b(\Delta_p)$, for all conditions $p \in P$.
\end{itemize}
\end{example}
The syntactic forcing described in the example above is used to prove completeness. 
The constants from $C$ are traditionally called Henkin constants and are used as witnesses for existentially quantified sentences obtained by extending an initial theory to a maximally consistent set of sentences.

As usual, forcing properties determine suitable relations between conditions and sentences.
\begin{definition} [Forcing relation]
Let $\mathbb P=\langle P,\leq,\Delta,f \rangle$ be a forcing property.
\emph{The forcing relation} $\Vdash$ between conditions $p\in P$ and sentences from $\Sen(\Delta_p)$ is defined by induction on the structure of sentences, as follows:
\begin{itemize}
\item $p\Vdash \varphi$ if $\varphi \in f(p)$, for all atomic sentences $\varphi\in\Sen_b(\Delta_p)$.
		
\item  $p\Vdash t_1\stackrel{\act_1\comp\act_2}\Longrightarrow  t_2$ if $p\Vdash t_1 \stackrel{\act_1}\Longrightarrow t $ and $p\Vdash t \stackrel{\act_2}\Longrightarrow t_2$ for some $t\in T_{\Delta_p}$.
		
\item $p\Vdash t_1\stackrel{\act_1\cup\act_2}\Longrightarrow  t_2$ if $p\Vdash t_1 \stackrel{\act_1}\Longrightarrow t_2 $ or $p\Vdash t_1 \stackrel{\act_2}\Longrightarrow t_2$.
		
\item $p\Vdash t_1\stackrel{\act^*}\Longrightarrow  t_2$ if $p\Vdash t_1\stackrel{\act^n}\Longrightarrow  t_2$ for some natural number $n\in\omega$.
		
\item $p\Vdash \neg \phi$ if there is no $q\geq p$ such that $q\Vdash \phi$. 
		
\item $p\Vdash \vee \Phi$ if $p\Vdash \phi$ for some $\phi\in \Phi$.
		
\item $p\Vdash \Exists{X}\phi$ if $p\Vdash \theta(\phi)$ for some substitution $\theta:X \to T_{\Delta_p}$.
\end{itemize}
The relation $p\Vdash \phi$ in $\mathbb{P}$, is read as $p$ forces $\phi$.
We say that $p$ weakly forces $\phi$, in symbols, $p\Vdash^w \phi$, if $p\Vdash\neg\neg\phi$.
\end{definition}
A few basic properties of forcing are presented below.
\begin{lemma}[Forcing properties] \label{lemma:fp}
Let $\mathbb P=(P,\leq,\Delta,f)$ be a forcing property. 
For all conditions $p\in P$ and all sentences $\phi\in\Sen(\Delta_p)$ we have: 
\begin{enumerate} [topsep=3pt]
\item \label{fp-1} $p\Vdash \neg\neg \phi$ iff for each $q\geq p$ there is a condition $r\geq q$ such that $r\Vdash \phi$.
		
\item \label{fp-2} If $p\leq q$ and $p\Vdash \phi$ then $q\Vdash \phi$.
		
\item  \label{fp-3} If $p\Vdash \phi$ then $p\Vdash \neg\neg \phi$.
		
\item \label{fp-4} We can not have both $p\Vdash \phi$ and $p\Vdash \neg \phi$.
\end{enumerate}
\end{lemma}
The second property stated in the above lemma shows that the forcing relation is preserved along inclusions of conditions.
The fourth property shows that the forcing relation is consistent, that is, a condition cannot force all sentences.
The remaining conditions are about negation.
\begin{definition}[Generic set] \label{def:gs}
Let $\mathbb P=(P,\leq,\Delta,f)$ be a forcing property.
A subset of conditions $G\subseteq P$ is generic if
\begin{enumerate}
\item\label{gs-1} $G$ is an ideal, that is:
\begin{enumerate*}
\item for all $p\in G$ and all $q\leq p$ we have $q\in G$, and
		
\item for all $p,q\in G$ there exists $r\in G$ such that $p\leq r$ and $q\leq r$; and
\end{enumerate*}		
\item\label{gs-2} for all conditions $p\in G$ and all sentences $\phi\in \Sen(\Delta_p)$ there exists a condition $q\in G$ such that $q\geq p$ and either $q\Vdash \phi$ or $q\Vdash \neg \phi$ holds.
\end{enumerate}
We write $G\Vdash \phi$ if $p\Vdash \phi$ for some $p\in G$.
\end{definition}
A generic set $G$ describes a reachable model which satisfies all sentences forced by the conditions in $G$.

\begin{lemma}[Existence]\label{lemma:gsl}
Let $\mathbb P=(P,\leq,\Delta,f)$ be a forcing property.
If any signature in $\{\Delta_p\}_{p\in P}$ is countable then every $p\in P$ belongs to a generic set.
\end{lemma}
\begin{proof}[Proof sketch]
Let $\text{pair}:\omega\times\omega\to\omega$ be a bijective function defined by $\text{pair}(i,j):=\bigl((i+j)(i+j+1)+2j\bigr)/2$ for all $i,j\in\omega$.
For all conditions $p\in P$, let $\psi_p:\omega\to\Sen(\Delta_p)$ be a bijective mapping, which gives an enumeration of $\Sen(\Delta_p)$.
We define an increasing chain of conditions $p_{0} \leq p_{1} \leq\dots$ in $P$ recursively.
Let $p=p_0$. For the induction step, we assume that we have defined $p_n$ and we define $p_{n+1}$.
Notice that there are unique natural numbers $i,j\in\omega$ such that $n=\text{pair}(i,j)$ and $i,j\leq n$.
\begin{itemize}[topsep=3pt]
\item If there is $q\geq p_n$ such that $q\Vdash \psi_{p_i}(j)$, then let $p_{n+1}\coloneqq q$. 
\item Otherwise, $p_{n+1}:=p_n$, which means that $p_{n+1}\Vdash \neg\psi_{p_i}(j)$.
\end{itemize}
Then $G\coloneqq \{q \in P \mid q \leq p_{n} \text{ for some } n \in \omega \}$ is generic and contains $p$.
\end{proof}
Lemma~\ref{lemma:gsl} is the key for a modular approach to forcing and it is the equivalent of the Lindenbaum's lemma from Henkin's method for proving completeness.
\begin{remark}\label{rem:gsr}
Since $\Delta:(G,\leq)\to \Sig$ is a directed diagram of signature inclusions, one can construct a co-limit  $\mu: \Delta \Rightarrow \Delta_G$ of the functor $\Delta:(G,\leq)\to \Sig$ such that $\mu_p:\Delta_p\to\Delta_G$ is an inclusion for all $p\in G$.
\end{remark}
The results which leads to completeness are developed over the signature~$\Delta_G$.
If $\mathbb{P}$ is the syntactic forcing described in Example~\ref{ex:syntactic-forcing}, then $\Delta_G$ is obtain from the base signature $\Sigma$ by adding all Henkin constants from $\{\Delta_p\}_{p\in G}$.
In general, $\Delta_G$ does not contain all Henkin constants from $C$, which is one of the major differences between the classical approach and the present developments.
\begin{definition} [Generic model]
Let $\mathbb P=(P,\leq,\Delta,f)$ be a forcing property and $G\subseteq P$ a generic set.
A model $\A$ defined over $\Delta_G$ is a \emph{generic model} for $G$ iff for every sentence $\phi\in \bigcup_{p\in G}\Sen(\Delta_p)$, we have $\A\models \phi \mbox{ iff } G\Vdash \phi$.
\end{definition}
The notion of generic model is the semantic counterpart of the definition of generic set.
The following result shows that every generic set has a generic model.
\begin{theorem}[Generic Model Theorem] \label{th:gm}
Let $\mathbb P=(P,\leq,\Delta,f)$ be a forcing property and $G\subseteq P$ a generic set. 
Then there is a generic model $\A$ for $G$ which is countable and reachable.
\end{theorem}
\begin{proof}[Proof sketch]
We define the set of all atomic sentences $B\coloneqq \{ \phi \in \Sen_b(\Delta_G) \mid G\Vdash \phi \}$ forced by the generic set $G$.
The basic model $\A^B$ given by Lemma~\ref{lemma:basic} is the generic model for $G$.
\end{proof}
\section{Completeness} \label{sec:completeness}
The logical framework in which the results are developed in this section is the fragment $\TA^c$ obtained from $\TA$ by restricting the syntax to at most countable signatures.
The syntactic forcing property defined in Example~\ref{ex:syntactic-forcing} is the starting point for proving completeness.
Therefore, throughout this section, we let $\mathbb P=(P,\leq,\Delta,f)$ be a syntactic forcing property in $\TA^c$ as described in Example~\ref{ex:syntactic-forcing}.
In particular, all signatures in $\{\Delta\}_{p\in P}$ are at most countable.
For the sake of simplicity, we write $p\vdash \phi$ iff $\Gamma_p\vdash_{\Delta_p} \phi$, for all conditions $p=(\Delta_p,\Gamma_p)$~in~$P$.
\begin{theorem} \label{th:sfp}
For all $p \in P$ and all $\phi\in\Sen(\Delta_p)$, we have
$p\Vdash^{w} \phi \text{ iff } p\vdash \phi$.
\end{theorem}
The above theorem says that a sentence is entailed by a condition if and only if it is weakly forced by that condition.
In other words, the entailment relation is the weak forcing relation.
Now, we can interpret Lemma~\ref{lemma:gsl} in the present context given by the syntactic forcing property $\mathbb{P}$ set above.
The following result is a direct consequence of Theorem~\ref{th:sfp} and Lemma~\ref{lemma:gsl}.

\begin{corollary} [Lindenbaum's lemma] \label{cor:lind}
Assume the following: 
\begin{itemize}[topsep=3pt]
\item a condition $p^\circ=(\Delta_{p^\circ},\Gamma_{p^\circ})$ from $P$, and
\item a generic set $G$ which contains $p^\circ$ (by Lemma~\ref{lemma:gsl}).
\end{itemize}
Let $\Delta_G$ be the vertex of the co-limit $\mu:\Delta\to \Delta_G$ of the functor $\Delta:(G,\leq)\to \Sig$ defined in Remark~\ref{rem:gsr}.
Then $\Gamma_G=\bigcup_{p\in G} \Gamma_p$ is a maximally consistent set which includes $\Gamma_{p^\circ}$.
\end{corollary}
The following example shows that $\Delta_G$ does not contain all Henkin constants defined for the base signature.
\begin{example}
Let $\Sigma$ be the base signature defined as follows:
\begin{enumerate*}[label=(\alph*)]
\item $S\coloneqq\{s_i\mid i\in \omega\}$,
\item $F\coloneqq\{c:\to s_0,d:\to s_0\}$
\item $M\coloneqq\emptyset$, and
\item $P\coloneqq\{\lambda\}$.
\end{enumerate*}
Let $\Gamma$ be the set of sentences over $\Sigma$ which consists of:
\begin{enumerate*}[label=(\alph*)]
\item $ c \stackrel{\lambda^*}\Longrightarrow d$, and
\item $(\Exists{x_n}\top) \rightarrow \neg(c \stackrel{\lambda^n}\Longrightarrow d) $ for all $n\in\omega$, where $x_n$ is a variable of sort $s_n$.
\end{enumerate*}
\end{example}
The first sentence says that there is a transition from $c$ to $d$ in a finite number of steps.
For each natural number $n$, the sentence $(\Exists{x_n}\top) \rightarrow \neg(c \stackrel{\lambda^n}\Longrightarrow d)$ says that if the sort $s_n$ is not empty then there is no transition from $c$ to $d$ in exactly $n$ steps. 
Recall that $C=\{C_{s_n}\}_{n\in\omega}$ is the set of all Henkin constants, and $\card(C_{s_n})=\omega$ for all natural numbers $n$. 
Notice that $p^\circ=(\Sigma,\Gamma)$ is consistent, but $q=(\Sigma[C],\Gamma)$ is not consistent.
By Corollary~\ref{cor:lind}, one can extend $\Gamma$ to a maximally consistent set of sentences $\Gamma_G$.
Unlike in classical first-order logic, $\Gamma_G$ does not contain all Henkin constants from $C$.

\begin{theorem} [Downwards L\"owenheim-Skolem Theorem]\label{th:LST}
For	any consistent set of sentences $\Gamma$ defined over a countable signature $\Sigma$, there exists a countable $\Sigma$-model $\mathfrak{A}$ that satisfies $\Gamma$.
\end{theorem}
\begin{proof}
Let $\mathbb{P}=(P,\leq,\Delta,f)$ be the forcing property described in Definition~\ref{def:forcing}.
Notice that $p\coloneqq (\Sigma,\Gamma)$ is a condition from $P$.
Since all signatures are countable, by Lemma~\ref{lemma:gsl}, $p$ belongs to a generic set $G$. 
By Theorem~\ref{th:gm}, $G$ has a generic model $\B$ which is countable and reachable. 
In particular, $\B\models_{\Delta_G}\Gamma$. 
Let $\A\coloneqq\B\red_\Sigma$, and by the satisfaction condition, $\A\models_{\Sigma}\Gamma$. 
\end{proof}

\begin{theorem}[Completeness]
For all sets of sentences $\Gamma$ and all sentences $\phi$ defined over a countable signature $\Sigma$, we have:
$\Gamma\vdash_\Sigma\phi$ iff $\Gamma\models_\Sigma\phi$.
\end{theorem}
\begin{proof}
The forward implication holds because all proof rules are sound.
For the backwards implication  assume $\Gamma\not\vdash_\Sigma \phi$.
We have $\Gamma\cup \{\neg\phi\}\not\vdash_\Sigma \bot$.
By Theorem~\ref{th:LST}, there is a countable $\Sigma$-model $\A$ such that $\A\models_\Sigma\Gamma\cup\{\neg\phi\}$.
Therefore, $\Gamma\not\models_\Sigma\phi$.
\end{proof}
\section{Conclusions}
In this study, we have defined an extension of many-sorted first-order logic, called transition algebra, that offers explicit support for state transitions; 
furthermore, we have investigated its logical properties in order to apply the institutional model theory approach to new algebraic specification languages based on this logic, and with a greater expressivity than Maude and CafeOBJ.
Transition algebra satisfies desirable properties such as truth invariance under change of signature, and has an expressive power that goes beyond that of ordinary first-order logic, which is important for formal-verification purposes. Our efforts have focused on two main aspects of transition algebra: 
first, on its formal-specification capabilities, i.e., to show that it forms a proper extension of first-order equational logic; 
and second, on support for formal verification, for which we have studied a number of model-theoretic properties, syntactic entailment and, most importantly, soundness and completeness results.

Concerning its formal-specification capabilities, transition algebra blends features of dynamic logic with features of many-sorted first-order logic.
From the former, it borrows the idea of expressing the dynamics of a system by means of actions, which are built from atomic transitions using composition, iteration, and so on.
The iteration of an action is a key feature because it allows us to express reachability, which is not possible in ordinary first-order logic.
From the latter, our logic borrows term-building operators and quantifiers. This allows us to capture system states as terms, and hence to reason about the structure of states more freely and in a more complex manner than it is possible in dynamic logic.

For verification purposes, our contribution is twofold: on one hand, we have introduced a sound proof system for transition algebra; and on the other hand, we have developed a new general method for proving completeness based on forcing.
The latter is highly important, because it has enabled us to circumvent the lack of compactness of transition algebra, which prevents the use of readily available methods for proving completeness.
Moreover, it also overcomes a significant limitation of existing forcing techniques, namely their reliance on models with non-empty carriers, which is another basic property (like compactness) that does not hold for transition algebra.
We have demonstrated the use of this extended forcing technique to show that the proof system for transition algebra is complete.
We aim to further develop and apply this technique to extensions of transition algebra that take into account, for example, subsorting -- to which we have already alluded in this paper.
Furthermore, future research includes applying forcing to prove omitting types theorem for logical systems that interpret sorts as sets, possibly empty, thus upgrading the results from  \cite{gai-ott,GainaBK23}.
Subsequently, the application of omitting types theorem to Robinson consistency property and interpolation, as demonstrated in  \cite{BadiaKG22}, remains a feasible avenue for exploration.

\bibliographystyle{plainurl}
\bibliography{ta}
\appendix
\section{Proofs for the results presented in Section~\ref{section:transition-algebra}}

In order to prove Proposition~\ref{proposition:trans_models} the following lemma is needed.
\begin{lemma} \label{lemme:termactchi}
For all signature morphisms $\chi:\Sigma\to\Sigma'$,
all $\Sigma'$-models $\A$,
all terms $t\in T_\Sigma$ and
all actions $\act\in A$,
we have
\begin{enumerate}
	\item \label{lemme:termchi} $\chi(t)^{\A}=t^{(\A\upharpoonright\chi)}$ and 
	\item \label{lemme:termact} $\chi(\act)_{\chi(s)}^\A=\act_s^{(\A\upharpoonright\chi)}$ for all sorts $s\in S$.
\end{enumerate}

\end{lemma}
\begin{proof}
	We prove the first statement by induction on the structure of terms.
	\begin{proofcases}
		\item [$c:\to s\in F$]
		From the definition of reduct, we get $\chi(c)^{\A}=c^{(\A\upharpoonright\chi)}$.
		\item [$\sigma(t_{1},\dots,t_{n})$]
		By induction hypothesis, we have $\chi(t_{i})^{\A}=t_{i}^{\A\upharpoonright\chi}$ for all $i\in\{1,\dots,n\}$.
It follows that
$\chi(\sigma(t_{1},\dots,t_{n}))^{\A}= 
\chi(\sigma)^{\A}(\chi(t_{1})^{\A},\dots,\chi(t_{n})^{\A})=
\sigma^{(\A\upharpoonright_{\chi})}(t_{1}^{(\A\upharpoonright_{\chi})},\dots,t_{n}^{(\A\upharpoonright_{\chi})})=
\sigma(t_{1},\dots,t_{n})^{(\A\upharpoonright\chi)}$.
\end{proofcases}

For the second statement, it suffices to prove that 
$d_1\stackrel{\act}\Longrightarrow d_2$ in $\A\red_\chi$ iff 
$d_1\stackrel{\chi(\act)}\Longrightarrow d_2$ in $\A$.
We proceed by induction on the structure of actions:
\begin{proofcases}
	\item [$\lambda\in L$]
	Straightforward, by the definition of $\lambda^{\A\red_\chi}$.
	\item [$\act_{1}\comp\act_{2}$]
	$d_1\stackrel{\act_1\comp\act_2}\Longrightarrow d_2$ in $\A\red_\chi$ iff
	$d_1\stackrel{\act_1}\Longrightarrow d$ and $d\stackrel{\act_2}\Longrightarrow d_2$ for some element $d$ in $\A\red_\chi$ iff (by IH)
	$d_1\stackrel{\chi(\act_1)}\Longrightarrow d$ and $d\stackrel{\chi(\act_2)}\Longrightarrow d_2$ for some element $d$ in $\A$ iff
	$d_1\stackrel{\chi(\act_1\comp\act_2)}\Longrightarrow d_2$ in $\A$.
	
	\item [$\act_{1}\cup\act_{2}$] This case is similar to the one above.
	
	\item [$\act^*$]	
	$d_1\stackrel{\act^*}\Longrightarrow d_2$ in $\A\red_\chi$ iff
	$d_1\stackrel{\act^n}\Longrightarrow d_2$ in $\A\red_\chi$ for some natural number $n\in\omega$ iff (by IH)
	$d_1\stackrel{\chi(\act)^n}\Longrightarrow d_2$ in $\A$ for some natural number $n\in\omega$ iff
	$d_1\stackrel{\chi(\act)^*}\Longrightarrow d_2$ in $\A$.
\end{proofcases}
\end{proof}
The proof of the satisfaction condition is given below.
\begin{proof}[Proof of Proposition~\ref{proposition:trans_models}]
We show this by structural induction on $\phi$.
	\begin{proofcases}[itemsep=1ex]
		\item [$t_{1}=t_{2}$]
		
		$\A\models\chi(t_{1}=t_{2})$ iff 
		$\A\models\chi(t_{1})=\chi(t_{2})$ iff 
		$\chi(t_{1})^{\A}=\chi(t_{2})^{\A}$ iff (by Lemma~\ref{lemme:termactchi} (\ref{lemme:termchi}))
		$t_{1}^{\A\upharpoonright_{\chi}}=t_{2}^{\A\upharpoonright_{\chi}}$ iff 
		$\A\upharpoonright_{\chi}\models t_{1}=t_{2}$.
		
		\item [$t_1 \stackrel{\act}\Rightarrow t_2$] 
		$\A\models \chi(t_1\stackrel{\act}\Rightarrow t_2)$ iff
		$\A\models \chi(t_1)\stackrel{\chi(\act)}\Longrightarrow \chi(t_2)$ iff
		$\chi(t_1)^\A \stackrel{\chi(\act)}\Longrightarrow \chi(t_2)^\A$ iff (by Lemma~\ref{lemme:termactchi} (\ref{lemme:termact}))
		$t_1^{\A\red_\chi} \stackrel{\act}\Longrightarrow  t_2^{\A\red_\chi}$ iff 
		$\A\upharpoonright_{\chi}\models t_1\stackrel{\act}\Rightarrow t_2$.
		
		\item [$\neg\phi$] 
		$\A\models\chi(\neg\phi)$ iff 
		$\A\models\neg\chi(\phi)$ iff 
		$\A\not\models\chi(\phi)$ iff (by IH)
		$\A\upharpoonright_{\chi}\not\models\phi$ iff
		$\A\upharpoonright_{\chi}\models\neg\phi$.
		
\item [$\vee\Phi$] 
$\A\models\chi(\vee\Phi)$ iff 
$\A\models\vee\chi(\Phi)$ iff 
$\A\models\chi(\phi)$ for some $\phi\in\Phi$ iff (by IH)
$\A\upharpoonright_{\chi}\models\phi$ for some $\phi\in\Phi$ iff 
$\A\upharpoonright_{\chi}\models\vee\Phi$.
		
\item [$\Exists{X}\phi$] The following are equivalent:

$\A\models\chi(\Exists{X}\phi)$ iff 
$\A\models\Exists{X'}\chi'(\phi)$ iff
		
$\mathfrak{D}\models\chi'(\phi)$ for some $\Sigma'[X']$-expansion $\D$ of $\A$ iff  (by IH)

\begin{center}
		\begin{tikzcd} 
\Sigma{[X]} \ar[r,"\chi'"] & \Sigma'{[X']}\\
\Sigma \ar[u,hook] \ar[r,"\chi"] & \Sigma' \ar[u,hook,swap]
\end{tikzcd}
\end{center}
		
$\mathfrak{D}\upharpoonright_{\chi'}\models\phi$ for some $\Sigma'[X']$-expansion $\D$ of $\A$ iff 
		
$\B\models\phi$ for some $\Sigma[X]$-expansion $\B$ of $\A\red_\chi$ iff
		
$\A\upharpoonright_{\chi}\models\Exists{X}\phi$.
\end{proofcases}
\end{proof}
\section{Proofs for the results presented in Section~\ref{sec:entailments}}
\begin{proof}[Proof of Lemma~\ref{lemma:basic-compact} (Basic compactness)]
Let $\vdash^\omega$ be the compact entailment relation determined by $\vdash^b$:
$\Gamma\vdash^\omega\Phi$ iff for each $\Phi'\subseteq \Phi$ finite there exists $\Gamma'\subseteq \Gamma$ finite such that $\Gamma'\vdash^b \Phi'$.
One can easily show that $\vdash^\omega$ is an entailment relation and it is closed under the basic proof rules enumerated in Definition~\ref{def:basic}. 
Since $\vdash^b$ is the least entailment relation closed under the basic proof rules, we get $\vdash^b=\vdash^\omega$.
\end{proof}

\begin{proof}[Proof of Lemma~\ref{lemma:basic}]
	Let $\Gamma$ be a set of atomic sentences and create a basic model of $\Gamma$. 
	First, define $\equiv_{\Gamma}$ by
	\begin{center}
		$\equiv_{\Gamma,s}\coloneqq\{\, \pos{t,u} \mid t,u\in T_{\Sigma,s}\mbox{ and }\Gamma\models t=u\}$
		for all sorts $s\in S$.
	\end{center}
One can easily show that $\equiv_{\Gamma}=\{\equiv_{\Gamma,s}\}_{s\in S}$ is a congruence on $T_\Sigma$.
Let $\A^\Gamma$ be the model obtained from the algebra $T_\Sigma/_{\equiv_\Gamma}$ by interpreting each symbol $\lambda\in L$ by $\lambda^{(\A^\Gamma)}=\{\pos{[t],[u]} \mid \Gamma\models t\stackrel{\lambda}\Rightarrow u\}$, where $[t]=t/_{\equiv_\Gamma}$ is the equivalence class of $t$ under $\equiv_\Gamma$.	
We show that for all $f:s_1\dots s_n\to s\in M$, the function $f^{(\A^\Gamma)}$ is monotone.
	\begin{itemize}
		\item []
		Assume that $[t_j]\stackrel{\lambda}\Longrightarrow [u_j]$. 
		By the definition of $\lambda^{(\A^\Gamma)}$, 
		$\Gamma\models t_j\stackrel{\lambda}\Rightarrow u_j$.
It follows that	$\Gamma\models f(t_1,\dots,t_j,\dots,t_n) \stackrel{\lambda} \Rightarrow f(t_1,\dots,u_j,\dots,t_n)$.
By the definition of $\lambda^{(\A^\Gamma)}$, we have that
$f^{(\A^\Gamma)}([t_1],\dots,[t_j],\dots,[t_n]) \stackrel{\lambda}\Longrightarrow f^{(\A^\Gamma)}([t_1],\dots,[u_j],\dots,[t_n])$.
\end{itemize}

Hence, $\A^\Gamma$ is well-defined.
By the definition of $\A^\Gamma$, for all $\varphi\in\Sen_b(\Sigma)$ we have that:
$\A^\Gamma \models\varphi \text{ iff } \Gamma\models\varphi$.

Next we show that $\A^\Gamma$ is a basic model, i.e. for all models $\A$, we have that
$\A\models\Gamma \text{ iff there is a unique homomorphism } h:\A^{\Gamma}\to\A$.
	\begin{proofcases}
		\item [$\Leftarrow$] For the backwards implication, assume a homomorphism $h:\A^{\Gamma}\to\A$; 
		since homomorphisms preserves the satisfaction of atomic sentences and $\A^\Gamma\models \Gamma$, we obtain $\A\models\Gamma$.
		\item [$\Rightarrow$]
		We define a homomorphism $h:\A^\Gamma\to \A$ by $h([t])=t^\A$ for all terms $t\in T_\Sigma$.
\begin{itemize}
\item We show that $h$ is a function, that is,  $[t_1]=[t_2]$ implies $h([t_1])=h([t_2])$ for all terms $t_1,t_2\in T_\Sigma$.
Indeed, if $[t_1]=[t_2]$ then $t_1\equiv_\Gamma t_2$, which means $\Gamma\models t_1=t_2$.
Since $\A\models \Gamma$, we have $\A\models t_1=t_2$, which means $t_1^\A=t_2^\A$.
Hence, $h([t_1])=h([t_2])$.
			
\item We show that $h$ is homomorphic, that is, $h(\sigma^{(\A^\Gamma)}([t_1],\dots,[t_n]))=\sigma^\A(h([t_1]),\dots,h(t_n))$ for any function symbol $\sigma:s_1\dots s_n\to s\in F$ and any term $t_i\in T_{\Sigma,s_i}$, where $i\in\{1,\dots,n\}$.
Indeed, 
\begin{align*} 
h(\sigma^{(\A^\Gamma)}([t_1],\dots [t_n]))= \\
h([\sigma(t_1,\dots,t_2)])= \\
\sigma(t_1,\dots,t_n)^\A= \\
\sigma^\A(t_1^\A,\dots,t_n^\A)= \\
\sigma^\A(h([t_1]),\dots,h([t_n])).
\end{align*}

\item We show that $h$ preserves the validity of relations, that is, if $[t_1]\stackrel{\lambda}\Longrightarrow [t_2]$ then $t_1^\A\stackrel{\lambda}{\Longrightarrow}t_2^\A$ for all $\lambda \in L$ and all terms $t_1,t_2\in T_{\Sigma,s}$. 
Indeed, if $[t_1]\stackrel{\lambda}\Longrightarrow [t_2]$ then $\Gamma\models t_1\stackrel{\lambda}\Rightarrow t_2$.
Since $\A\models \Gamma$, we have $\A\models t_1\stackrel{\lambda}\Rightarrow t_2$, which means that $t_1^\A\stackrel{\lambda}\Longrightarrow t_2^\A$.
			
\item  Since the elements of $\A^\Gamma$ are interpretations of terms, $h:\A^\Gamma\to \A$ is unique.
\end{itemize}
\end{proofcases}
\end{proof}

\begin{proof}[Proof of Proposition~\ref{prop:basic-complete} (Basic completeness)]
First, we define the following congruence $\equiv_{\Gamma}$ on $T_\Sigma$:
\begin{center}
$\equiv_{\Gamma,s}\coloneqq\{\, \pos{t,u} \mid t,u\in T_{\Sigma,s}\mbox{ and }\Gamma\vdash^{b} t=u\}$ for all sorts $s\in S$.
\end{center}
By the basic proof rules,
$\equiv_{\Gamma}=\{\equiv_{\Gamma,s}\}_{s\in S}$ is well-defined.
Let $\D^\Gamma$ be the model obtained from the algebra $T_\Sigma/_{\equiv_\Gamma}$ by interpreting each symbol $\lambda\in L$ by $\lambda^{(\D^\Gamma)}=\{\pos{[t],[u]} \mid \Gamma\vdash^{b} t\stackrel{\lambda}\Rightarrow u\}$, where $[t]=t/_{\equiv_\Gamma}$ is the equivalence class of $t$ under $\equiv_\Gamma$.
By $(P)$ from Definition~\ref{def:basic},
	$\lambda^{(\D^\Gamma)}$ is well-defined.
We show that for all $f:s_1\dots s_n\to s\in M$, the function $f^{(\D^\Gamma)}$ is monotone:
	\begin{itemize}
		\item []
		Assume that $[t_j]\stackrel{\lambda}\Longrightarrow [u_j]$. 
		By the definition of $\lambda^{(\D^\Gamma)}$, we have 
		$\Gamma\vdash^{b} t_j\stackrel{\lambda}\Rightarrow u_j$.
		
		By $(\mathsf{M})$,
		$\Gamma\vdash^{b} f(t_1,\dots,t_j,\dots,t_n) \stackrel{\lambda} \Rightarrow f(t_1,\dots,u_j,\dots,t_n)$.
		
		By the definition of $\lambda^{(\D^\Gamma)}$,
		$f^{(\D^\Gamma)}([t_1],\dots,[t_j],\dots,[t_n]) \stackrel{\lambda}\Longrightarrow f^{(\D^\Gamma)}([t_1],\dots,[u_j],\dots,[t_n])$.
\end{itemize}
Hence, $\D^\Gamma$ is well-defined.
\
\begin{enumerate}
\item By the definition of $\D^\Gamma$, 
$\D^\Gamma \models\varphi$ iff $\Gamma\vdash^b\varphi$.
By $(Monotonicity)$, we get $\D^\Gamma\models\Gamma$. 
\item By soundness, $\Gamma\vdash^b\varphi$ implies $\Gamma\models^b\varphi$.
\item Since $\D^\Gamma\models \Gamma$, $\Gamma\models^b\varphi$ implies $\D^\Gamma\models\varphi$.
\end{enumerate}
Hence,
$\Gamma \models^b\varphi$ iff $\Gamma\vdash^b\varphi$ iff $\D^\Gamma\models\varphi$.
\end{proof}
\begin{proof}[Proof of Proposition~\ref{prop:omega-compact} ($\omega_1$-compactness)]
For the first statement, we define $\vdash^{\omega_1}$, the $\omega_1$-compact entailment relation determined by $\vdash$:
$\Gamma\vdash^{\omega_1}\Phi$ iff for each at most countable subset $\Phi'\subseteq \Phi$ there exists $\Gamma'\subseteq \Gamma$ at most countable such that $\Gamma'\vdash \Phi'$.
One can easily show that $\vdash^{\omega_1}$ is an entailment relation and it is closed under the dynamic proof rules enumerated in Definition~\ref{def:dynamic-rules}. 
Since $\vdash$ is the least entailment relation closed under the dynamic proof rules, we get $\vdash=\vdash^{\omega_1}$.

For the second statement, let $\Sigma=(S,F\supseteq M,L)$ be an uncountable signature with at least one transition label $\lambda$.
It is sufficient to define a set of sentences $\Gamma\subseteq\Sen(\Sigma)$ without a model such that any countable subset of $\Gamma$ has a model.
We consider three cases.
\begin{proofcases}
\item[$S$ is uncountable]
Let  $\{s_\alpha \mid \alpha<\card(S)\}$ be an enumeration of $S$.

In first-order logic, for any natural number $n$ and sort $s_\alpha$, we can construct a sentence expressing that there are at least $n$ elements of sort $s_\alpha$ using $n$ variables $X=\{x_1,\dots,x_n\}$ of sort $s_\alpha$:
		\[
		\varphi_\alpha^n:=\Exists{X}\bigwedge_{i\neq j} \neg(x_i=x_j)
		\]
In transition algebra, one can express that the number of elements of sort $s_\alpha$ is finite: 
\begin{center}
$\phi\,_\alpha^\omega \coloneqq (\Forall{z} \exists! x \cdot x\stackrel{\lambda}\Longrightarrow z \wedge \exists! y \cdot z\stackrel{\lambda}\Longrightarrow y)
\wedge  (\Forall{x,y} x\stackrel{\lambda^*}\Longrightarrow y) $
\end{center}
where, $\exists!z\cdot\varphi(z):=\Exists x\varphi(x)\wedge \Forall y\varphi(y) \Rightarrow x=y$, and the sorts of the variables are all $s_\alpha$.
		
The first term of the above conjunction means that every element has a unique successor and predecessor.
The second term of the above conjunction means that all elements are connected by the relation $\lambda$.
		
Let $\A$ be a model that satisfies $\phi_\alpha^\omega$. 
Suppose $\A_{s_\alpha}\neq\emptyset$. 
Take one element $a$ from $\A_{s_\alpha}$.
Since $a\stackrel{\lambda^*}\Longrightarrow a$ holds in $\A$, 
$a\stackrel{\lambda^n}\Longrightarrow a$ holds in $\A$ for some natural number n.
Therefore, there exists a sequence of elements $a_1,...,a_n$ from $\A_{s_\alpha}$ such that
\[
a=a_0 \stackrel{\lambda}\Longrightarrow a_1 \stackrel{\lambda}\Longrightarrow \cdots \stackrel{\lambda}\Longrightarrow a_{n-1} \stackrel{\lambda}\Longrightarrow a_{n}=a
\]
		Due to the uniqueness of successors and predecessors, all elements of $\A_{s_\alpha}$ appear in this sequence.
		Therefore, $\A_{s_\alpha}$ is finite.
		\\
		\\
		Using these sentences
		we define $\Gamma$ as:
		\[
		\Gamma:=
		\{ 
		\phi\,_\alpha^\omega \mid \alpha<\text{card}(S) \}
		\cup
		\{ \varphi_\alpha^n \Rightarrow \varphi_\beta^{n+1} \mid n\in\omega,\alpha<\beta<\text{card}(S) \}.
		\]
		What $\A\models\Gamma$ means is that all domains $\{\A_{s}\}_{s\in S}$ are finite and the order of their sizes is exactly the same as the order of the corresponding ordinals.
		This contradicts the fact that $S$ is uncountable.
		On the other hand, it is clear that any countable subset of $\Gamma$ has a model.
		\item[$F$ is uncountable] 
		Assume $S$ is at most countable.
		If not, this case is the same as the first one.
		For simplicity, we assume that $F$ has only constant symbols.
		Since $F$ is uncountable and $S$ is at most countable, for at least one sort $s\in S$, there are uncountably many constant symbols of sort $s$.
We define $\Gamma_s$ as follows:
		\[
		\Gamma_s:=\{ \neg(c=c') \mid (c:\to s), (c':\to s)\in F \text{ and } c\neq c' \}
		\]
One can express an at most countable number of elements for the sort $s$ as follows: 
\begin{center}
$\phi\,_{s}^{\omega_{1}} \coloneqq
(\Forall{z} \exists! x \cdot x\stackrel{\lambda}\Longrightarrow z \wedge \exists! y \cdot z\stackrel{\lambda}\Longrightarrow y)
\wedge 
\forall{x,y}\, (x\stackrel{\lambda^*}\Longrightarrow y \vee y\stackrel{\lambda^*}\Longrightarrow x)$
\end{center}
where the sort of each variable is $s$.
\footnote{This means that the upward L\"{o}wenheim–Skolem Theorem also does not hold in this logic.}
		
Let $\A$ be a model such that $\A\models \phi_s^{\omega_1}$. 
For all elements $a,b\in A_{s}$, we have 
$a \stackrel{\lambda^*}\Longrightarrow b$ or $b \stackrel{\lambda^*}\Longrightarrow a$, 
which means 
$a \stackrel{\lambda^n}\Longrightarrow b$ or $b \stackrel{\lambda^n}\Longrightarrow a$ for some natural number $n$.
By the uniqueness of successors and predecessors, all elements of $\A_{s}$ can be lined up in a sequence. Therefore, $\A_s$ is at most countable.
\\
\\
Let
\(
\Gamma:=\Gamma_s\cup\{\phi\,_{s}^{\omega_1}\}
\)
and assume $\A\models\Gamma$.
Since there are uncountably many constant symbols whose sort is $s$, by $\Gamma_s$ the size of the domain $\A_s$ is also uncountable, which contradicts $\A\models\phi\,_{s}^{\omega_1}$.
Therefore, there is no model that satisfies $\Gamma$. 
However any countable subset of $\Gamma$ has a model.
\item[$L$ is uncountable]
Let $s\in S$ be a sort and $\lambda\in L$ a transition label.
We define 
\begin{itemize}
\item $\Gamma_1:=\{ \Forall{x,y,z} 
(x\stackrel{\lambda_1}\Longrightarrow z \wedge x\stackrel{\lambda_2}\Longrightarrow y)
\Rightarrow\neg(y=z) \mid \lambda_1,\lambda_2\in P\setminus\{\lambda\} \text{ and } \lambda_1\neq\lambda_2\}$

which says that different relations with the same source have different targets, and
\item $\Gamma_2:=\{\Forall{x}\exists! y \cdot x\stackrel{\lambda'}\Longrightarrow y \mid \lambda'\in L\setminus\{\lambda\}\}$

which says that any relation is in fact a function.
\end{itemize}
Define $\Gamma:=\Gamma_1\cup\Gamma_2\cup\{\Exists{x}\top\}\cup\{\phi\,_{s}^{\omega_1}\}$, where the sort of $x$ is $s$, and $\phi\,_{s}^{\omega_1}$ says that the domain of $s$ is at most countable.
		Since there is an uncountable number of transition labels, by $\Gamma_1$, $\Gamma_2$ and $\Exists{x}\top$, any model of $\Gamma$ should have uncountable elements of sort $s$, which contradicts $\phi\,_{s}^{\omega_1}$.
Therefore, there is no model that satisfies $\Gamma$. 
However any countable subset of $\Gamma$ has a model.
\end{proofcases}
\end{proof}
\section{Proofs for the results presented in Section~\ref{sec:forcing}}
\begin{proof} [Proof of Lemma~\ref{lemma:fp} (Forcing properties)]\
	
	\begin{enumerate}
		\item $p \Vdash^{w} \varphi$ iff $p \Vdash \neg\neg\varphi$ iff for each $q \geq p$, $q \not\Vdash  \neg\varphi$ iff for each $q \geq p$, there exists $r \geq q$ such that $r \Vdash \varphi$.
		\item Suppose $p\leq q$ and $p\Vdash \varphi$. We show $q\Vdash \varphi$ by induction on sentence structure.
\begin{proofcases}
\item [$\varphi$ is atomic]
The conclusion follows from $f(p) \subseteq f(q)$.
			
\item [$t_1\stackrel{\act_1\comp\act_2}\Longrightarrow  t_2$]
Since $p \Vdash t_1\stackrel{\act_1\comp\act_2}\Longrightarrow  t_2$, $p \Vdash t_1\stackrel{\act_1}\Longrightarrow  t$ and $p \Vdash t\stackrel{\act_2}\Longrightarrow  t_2$ for some $t \in T_{\Delta_p}\subseteq T_{\Delta_q}$. 
By induction hypothesis, $q \Vdash t_1\stackrel{\act_1}\Longrightarrow  t$ and $q \Vdash t\stackrel{\act_2}\Longrightarrow  t_2$, which means $q \Vdash t_1\stackrel{\act_1\comp\act_2}\Longrightarrow  t_2$.
			
\item [$t_1\stackrel{\act_1\cup\act_2}\Longrightarrow  t_2$]
Since $p \Vdash t_1\stackrel{\act_1\cup\act_2}\Longrightarrow  t_2$, $p \Vdash t_1\stackrel{\act_1}\Longrightarrow  t_2$ or $p \Vdash t_1\stackrel{\act_2}\Longrightarrow  t_2$. 
By induction hypothesis $q \Vdash t_1\stackrel{\act_1}\Longrightarrow  t_2$ or $q \Vdash t_1\stackrel{\act_2}\Longrightarrow  t_2$, which means $q \Vdash t_1\stackrel{\act_1\cup\act_2}\Longrightarrow  t_2$.
			
\item [$t_1\stackrel{\act^*}\Longrightarrow  t_2$]
By induction hypothesis and the case corresponding to the composition of actions, the equivalence holds for $\act^n$, where $n\in\omega$, not only for $\act$.
Since $p \Vdash t_1\stackrel{\act^{*}}\Longrightarrow  t_2$ then, by definition, $p\Vdash t_1\stackrel{\act^n}\Longrightarrow  t_2$ for some natural number $n\in\omega$.
By induction hypothesis, $q\Vdash t_1\stackrel{\act^n}\Longrightarrow  t_2$. 
Hence, $q\Vdash t_1\stackrel{\act^*}\Longrightarrow  t_2$.	
\item [$\neg\varphi$]
It is clear from the definition of negation of forcing relation.
			
\item [$\vee\Phi$]
$p \Vdash \varphi$ for some $\varphi \in \Phi$. By induction hypothesis $q \Vdash \varphi$ which implies $q \Vdash \vee\Phi$.
			
\item [$\Exists{X}\varphi$]			
Since $p \Vdash \Exists{X}\varphi$ then $p \Vdash \theta(\varphi)$ for some substitution $\theta : X \to T_{\Delta_p}$.
By induction hypothesis, $q \Vdash \theta(\varphi)$.
By the definition of forcing relation, $q \Vdash \Exists{X}\varphi$.			
\end{proofcases}
\item Suppose $p \Vdash \varphi$. From \ref{fp-1}, it is sufficient to show that for each $q \geq p$, there exists $r \geq q$ such that $r \Vdash \varphi$. this is clear from \ref{fp-2}.
\item It is clear from the definition of negation.
\end{enumerate}
\end{proof}
\begin{proof}[Proof of Lemma~\ref{lemma:gsl} (Existence)]
Let $\text{pair}:\omega\times\omega\to\omega$ be a bijective function defined by $\text{pair}(i,j):=\bigl((i+j)(i+j+1)+2j\bigr)/2$ for all $i,j\in\omega$.
For all conditions $p\in P$, let $\psi_p:\omega\to\Sen(\Delta_p)$ be a bijective mapping, which gives an enumeration of $\Sen(\Delta_p)$.
We define an increasing chain of conditions $p_{0} \leq p_{1} \leq\dots$ in $P$ recursively.
Let $p=p_0$. For the induction step, we assume that we have defined $p_n$ and we define $p_{n+1}$.
Notice that there are unique natural numbers $i,j\in\omega$ such that $n=\text{pair}(i,j)$ and $i,j\leq n$.
\begin{itemize}[topsep=3pt]
\item If there is $q\geq p_n$ such that $q\Vdash \psi_{p_i}(j)$, then let $p_{n+1}\coloneqq q$. 
\item Otherwise, $p_{n+1}:=p_n$, which means that $p_{n+1}\Vdash \neg\psi_{p_i}(j)$.
\end{itemize}
Then we show that $G\coloneqq \{q \in P \mid q \leq p_{n} \text{ for some } n \in \omega \}$ is generic and contains $p$.
\begin{itemize}
		\item Suppose $q\in G$, $r\in P$, and $r \leq q$. Since $q\in G$ there is some $n\in\omega$ such that $q\leq p_{n}$. Since $r\leq q \leq p_{n}$, $r$ is also in $G$.
		\item Suppose $q,q'\in G$. Since $q,q'\in G$ there is some $m,n\in\omega$ such that $q\leq p_{m}$ and $q'\leq p_{n}$ hold. By the definition of $\{p_{n}\}_{n\in\omega}$, $p_{m}\leq p_{n}$ or $p_{n}\leq p_{m}$ is true. Let $r$ be the larger one.
		Then $r\in G$ and $q,q'\leq r$.
		\item
		Suppose $q\in G$ and $\phi\in\Sen(\Delta_q)$.
		Since $q\in G$ there is some $i\in\omega$ such that $q\leq p_{i}$, and $\phi\in\Sen(\Delta_{p_i})$.
		Since $\psi_{p_i}:\omega\to\Sen(\Delta_{p_i})$ is bijection, $\phi=\psi_{p_i}(j)$ for some $j\in\omega$.
		Let $n=\text{pair}(i,j)$.
		Then, by the definition of $\{p_k\}_{k\in\omega}$, we have $p_{n+1}\geq p_i$, and either $p_{n+1}\Vdash \psi_{p_i}(j)$ or $p_{n+1}\Vdash \neg\psi_{p_i}(j)$ holds.
		Let $r:=p_{n+1}$, then $r\in G$, $r\geq q$, and either $r\Vdash \phi$ or $r\Vdash \neg\phi$.
	\end{itemize}
By definition of $G$,  $p$ belongs to $G$.
Therefore, every $p\in P$ belongs to a generic set.
\end{proof}
\begin{proof}[Proof of Theorem~\ref{th:gm} (Generic Model Theorem)]
Let $B=\{ \varphi \in \Sen_{b}(\Delta_G) \mid G\Vdash \varphi \}$.
We prove that for each $\phi \in \bigcup_{p\in G}\Sen(\Delta_p)$, $\A^{B} \models \phi$ iff $G \Vdash \phi$, where $\A^B$ is the basic model of $B$.
We proceed by induction on the structure of sentences.
\begin{proofcases}
\item [$\varphi \in \Sen_b(\Delta_G)$]\,
\begin{proofcases}
\item [$\Leftarrow$]
Suppose $G\Vdash \varphi$ then $\varphi\in B$, which means $B\models \varphi$.
Since $\A^{B}\models \varphi'$ iff $B\models \varphi'$ for all $\varphi' \in \Sen_{b}(\Delta_G)$, we have $\A^{B}\models \varphi$.
\item [$\Rightarrow$] 
Suppose $\A^{B} \models \varphi$ then we have $B \models \varphi$. 
By Lemma~\ref{lemma:basic-compact}, $B' \models \varphi$ for some finite subset $B' \subseteq B$.
Since $(G,\leq)$ is directed and $B'$ is finite, there exists $p\in G$ such that $B' \subseteq f(p)$. 
Since $B' \models \varphi$, we obtain $f(p) \models \varphi$.
Since $G$ is generic, we have $G \Vdash \varphi$ or $G \Vdash \neg\varphi$.
\begin{itemize}
\item[] Suppose towards a contradiction that $G \Vdash \neg\varphi$.
Then there is $q\in G$ such that $q \Vdash \neg\varphi$.
Since $G$ is generic, there is $r\in G$ such that $r \geq p$ and $r\geq q$. 
Since $r \geq q$ and $q\Vdash \neg\varphi$, by Lemma~\ref{lemma:fp}(\ref{fp-2}), we have $r \Vdash \neg\varphi$.
Also, since $r\geq p$, we have $f(p) \subseteq f(r)$; 
since $f(p) \models \varphi$, we obtain $f(r) \models \varphi$; 
by the definition of forcing property, $r'\Vdash \varphi$ for some $r'\geq r$.
Since $r'\geq r$ and $r\Vdash\neg \varphi$, by Lemma~\ref{lemma:fp}(\ref{fp-2}), $r'\Vdash\neg\varphi$.
It follows that $r'\Vdash\varphi$ and $r'\Vdash\neg\varphi$, which is a contradiction with Lemma~\ref{lemma:fp}(\ref{fp-4}).
\end{itemize}
Hence, $G \Vdash \varphi$.
\end{proofcases}
		
\item [$t_1\stackrel{\act_1\comp\act_2}\Longrightarrow  t_2$]\,
\begin{proofcases}
\item [$\Leftarrow$] 
Suppose $G \Vdash t_1\stackrel{\act_1\comp\act_2}\Longrightarrow t_2$. 
This means $p \Vdash t_1\stackrel{\act_1}\Longrightarrow t$ and $p \Vdash t\stackrel{\act_2}\Longrightarrow t_2$ for some $p\in G$ and $t\in T_{\Delta_p}$.
			By induction hypothesis, $\A^{B} \models t_1\stackrel{\act_1}\Longrightarrow t$ and $\A^{B} \models t\stackrel{\act_2}\Longrightarrow t_2$.
			Therefore, $\A^{B} \models t_1\stackrel{\act_1\comp\act_2}\Longrightarrow t_2$.
			\item [$\Rightarrow$]
			Suppose $\A^{B} \models t_1\stackrel{\act_1\comp\act_2}\Longrightarrow t_2$.
			Since $\A^{B}$ is reachable,
			$\A^{B} \models t_1\stackrel{\act_1}\Longrightarrow t$ and $\A^{B} \models t\stackrel{\act_2}\Longrightarrow t_2$ for some $t\in T_{\Delta_G}$.
			By induction hypothesis, $G \Vdash t_1\stackrel{\act_1}\Longrightarrow t$ and $G \Vdash t\stackrel{\act_2}\Longrightarrow t_2$.
			In other words
			$p \Vdash t_1\stackrel{\act_1}\Longrightarrow t$ and $q \Vdash t\stackrel{\act_2}\Longrightarrow t_2$ for some $p,q\in G$.
			Since $G$ is generic there is $r\in G$ such that $r \geq p$ and $r\geq q$.
			By Lemma~\ref{lemma:fp}(\ref{fp-2}), $r \Vdash t_1\stackrel{\act_1}\Longrightarrow t$ and $r \Vdash t\stackrel{\act_2}\Longrightarrow t_2$.
			By the definition of forcing relation, $r\Vdash t_1\stackrel{\act_1\comp\act_2}\Longrightarrow t_2$.
			Therefore, $G \Vdash t_1\stackrel{\act_1\comp\act_2}\Longrightarrow t_2$.
		\end{proofcases}
		
		\item [$t_1\stackrel{\act_1\cup\act_2}\Longrightarrow  t_2$]\,
		\begin{proofcases}
			\item [$\Leftarrow$] 
			Suppose $G \Vdash t_1\stackrel{\act_1\cup\act_2}\Longrightarrow t_2$.
			This means $p \Vdash t_1\stackrel{\act_1}\Longrightarrow t_2$ or $p \Vdash t_1\stackrel{\act_2}\Longrightarrow t_2$ for some $p\in G$. 
			By induction hypothesis, $\A^{B} \models t_1\stackrel{\act_1}\Longrightarrow t_2$ or $\A^{B} \models t_1\stackrel{\act_2}\Longrightarrow t_2$.
			Therefore $\A^{B} \models t_1\stackrel{\act_1\cup\act_2}\Longrightarrow t_2$.
			\item [$\Rightarrow$]
			Suppose $\A^{B} \models t_1\stackrel{\act_1\cup\act_2}\Longrightarrow t_2$. 
			This is equivalent to $\A^{B} \models t_1\stackrel{\act_1}\Longrightarrow t_2$ or $\A^{B} \models t_1\stackrel{\act_2}\Longrightarrow t_2$.
			By induction hypothesis, 
			$G \Vdash t_1\stackrel{\act_1}\Longrightarrow t_2$ or $G \Vdash t_1\stackrel{\act_2}\Longrightarrow t_2$.
			So,
			$p \Vdash t_1\stackrel{\act_1}\Longrightarrow t_2$ or $q \Vdash t_1\stackrel{\act_2}\Longrightarrow t_2$ for some $p,q\in G$.
			Since $G$ is generic there is $r\in G$\, such that $r \geq p$ and $r \geq q$. By Lemma~\ref{lemma:fp}(\ref{fp-2}), $r \Vdash t_1\stackrel{\act_1}\Longrightarrow t_2$ or $r \Vdash t_1\stackrel{\act_2}\Longrightarrow t_2$.
			By the definition of forcing relation, $r \Vdash t_1\stackrel{\act_1\cup\act_2}\Longrightarrow t_2$.
			Therefore, $G \Vdash t_1\stackrel{\act_1\cup\act_2}\Longrightarrow t_2$.
		\end{proofcases}
		
		\item [$t_1\stackrel{\act^*}\Longrightarrow  t_2$]\,
		\begin{proofcases}
			\item [$\Leftarrow$]
			By induction hypothesis and the case corresponding to the composition of actions,
			the equivalence holds for $\act^{n}$, where $n\in\omega$, not only for $\act$.
			
			Suppose $G \Vdash t_1\stackrel{\act^{*}}\Longrightarrow t_2$,
			then $p \Vdash t_1\stackrel{\act^{*}}\Longrightarrow t_2$ for some $p\in G$.
			By definition, $p \Vdash t_1\stackrel{\act^{n}}\Longrightarrow t_2$ for some $n\in\omega$.
			By induction hypothesis, $\A^{B} \models t_1\stackrel{\act^{n}}\Longrightarrow t_2$.
			Therefore, $\A^{B} \models t_1\stackrel{\act^{*}}\Longrightarrow t_2$.
			
\item [$\Rightarrow$]
Suppose $\A^{B} \models t_1\stackrel{\act^{*}}\Longrightarrow t_2$. 
This is equivalent to $\A^{B} \models t_1\stackrel{\act^{n}}\Longrightarrow t_2$ for some $n\in\omega$.
By induction hypothesis, $G \Vdash t_1\stackrel{\act^{n}}\Longrightarrow t_2$ in the same way as above.
Therefore, $G \Vdash t_1\stackrel{\act^{*}}\Longrightarrow t_2$.
\end{proofcases}		
\item [$\neg\phi$]\,
\begin{proofcases}
\item [$\Leftarrow$]

Suppose towards a contradiction that $p\Vdash \phi$ and $q\Vdash \neg\phi$ for some $p,q\in G$.
Since $G$ is generic there is $r\in G$ such that $r\geq p$ and $r\geq q$. 
By Lemma~\ref{lemma:fp}(\ref{fp-2}), we have $r\Vdash \phi$ and $r\Vdash \neg\phi$.
By Lemma~\ref{lemma:fp}(\ref{fp-4}), this is a contradiction.
			
Suppose $G\Vdash \neg\phi$. 
By the proof above, $G\not\Vdash \phi$. By induction hypothesis, $\A^{B}\not\models \phi$. Therefore $\A^{B} \models \neg\phi$.
			
			\item [$\Rightarrow$]
			Suppose $\A^{B}\models \neg\phi$. This is equivalent to $\A^{B}\not\models \phi$. By induction hypothesis, $G\not\Vdash \phi$.
			Since $G$ is generic, $G\Vdash \phi$ or $G\Vdash \neg\phi$.
			Therefore $G \Vdash \neg\phi$.
		\end{proofcases}
		
		\item [$\vee\Phi$]\,
		
		\begin{proofcases}
			\item [$\Leftarrow$]
			Suppose $G\Vdash\vee\Phi$. 
			By the definitions, $p\Vdash \phi$ for some $p\in G$ and some $\phi\in\Phi$. 
			By induction hypothesis, $\A^{B}\models \phi$. 
			Therefore, $\A^{B}\models \vee\Phi$.
			\item [$\Rightarrow$]
			Suppose $\A^{B}\models\vee\Phi$. There is some $\phi\in\Phi$ such that $\A^{B}\models \phi$. 
			By induction hypothesis, $G\Vdash \phi$. 
			So, $p\Vdash \phi$ for some $p\in G$. 
			Therefore, $p\Vdash \vee\Phi$.
		\end{proofcases}
		
		\item [$\Exists{X}\phi$]\,
		\begin{proofcases}
			\item [$\Leftarrow$]
			Suppose $G\Vdash \Exists{X}\phi$.
			We have $p\Vdash \Exists{X}\phi$ for some $p\in G$.
			By the definition of forcing relation, we have $p\Vdash \theta(\phi)$  for some substitution $\theta:X\to T_{\Delta_p}$.
			By induction hypothesis, $\A^{B}\models \theta(\phi)$. 
			By semantics, $\A^{B}\models \Exists{X}\phi$.
			\item [$\Rightarrow$]
			Suppose $\A^{B}\models \Exists{X}\phi$.
			Since $\A^{B}$ is reachable, there is $\theta:X\to T_{\Delta_G}$ such that $\A^{B}\models \theta(\phi)$.
By induction hypothesis, $G\Vdash \theta(\phi)$.
Therefore, $p\Vdash \theta(\phi)$ for some $p\in G$.
And we have $p\Vdash \Exists{X}\phi$ and $G\Vdash \Exists{X}\phi$.
\end{proofcases}
		
	\end{proofcases}
\end{proof}
\section{Proof for the results presented in Section~\ref{sec:completeness}}
In order to prove Theorem~\ref{th:sfp}, we need two additional lemmas.
\begin{lemma} \label{lemma:vD=vD^w}
	For all $p\in P$ and all $\phi\in\Sen(\Delta_p)$, the following are equivalent:
	\begin{enumerate}
		\item $p\vdash \phi$.
		\item For all  $q\geq p$ there exists $r\geq q$ such that  $r\vdash \phi$.
	\end{enumerate}
\end{lemma}\begin{proof}\
	\begin{proofcases}
		\item[$\Rightarrow$] Suppose $p\vdash \phi$ and $q\geq p$. 
		We show that there is $r\geq q$ such that $r\vdash \phi$. 
		Since $\Delta_q\supseteq\Delta_p$ and $\Gamma_q\supseteq \Gamma_p$, we have $q\vdash \phi$. 
		So, let $r\coloneqq q$.
		\item[$\Leftarrow$]
		Suppose towards a contradiction that $p\not\vdash \phi$.
		By $(Neg_D)$, $p\not\vdash \neg\neg\phi$.
		By $(Neg_I)$, $\Gamma_p\cup\{\neg\phi\} \not\vdash_{\Delta_p} \bot$, which means that $q:=(\Delta_p,\Gamma_p\cup\{\neg\phi\})$ belongs to $P$.
		Since $q\geq p$,
		there is $r\geq q$ such that $r\vdash \phi$.
		On the other hand, since $r\geq q$ and $q\vdash \neg\phi$, we get $r\vdash \neg\phi$, which is a contradiction with the consistency of $r$.
		Hence, $p\vdash\phi$.
	\end{proofcases}
\end{proof}
\begin{lemma} \label{lemma:up}
	For any condition $p \in P$, the following hold:
	\begin{enumerate}
		\item  \label{lemma:up1} If $p\vdash \varphi$ and $\varphi$ is atomic, there is $q \geq p$ such that $\varphi\in f(q)$.
		\item  \label{lemma:up2} If $p\vdash t_1\stackrel{\act_1\comp\act_2}\Longrightarrow t_2$, there are $q \geq p$ and $t\in T_{\Delta_q}$ such that $q\vdash t_1\stackrel{\act_1}\Longrightarrow t$ and $q\vdash t\stackrel{\act_2}\Longrightarrow t_2$.
		\item  \label{lemma:up3} If $p\vdash t_1\stackrel{\act_1\cup\act_2}\Longrightarrow t_2$, there is $q \geq p$ such that $q\vdash t_1\stackrel{\act_1}\Longrightarrow t_2$ or $q\vdash t_1\stackrel{\act_2}\Longrightarrow t_2$.
		\item  \label{lemma:up4} If $p\vdash t_1\stackrel{\act^*}\Longrightarrow t_2$, there are $q \geq p$ and $n\in\omega$ such that $q\vdash t_1\stackrel{\act^n}\Longrightarrow t_2$.
		\item  \label{lemma:up5} If $p\vdash \vee\Phi$, there are $q \geq p$ and $\phi\in\Phi$ such that $q\vdash \phi$.
		\item  \label{lemma:up6} If $p\vdash \Exists{X}\phi$, there are $q \geq p$ and $\theta:X\to T_{\Delta_q}$ such that $q\vdash \theta(\phi)$.
	\end{enumerate}
\end{lemma}
\begin{proof} \
	
	\begin{enumerate}
		\item Suppose $p\vdash \varphi$ and $\varphi$ is atomic. 
		Since $p\not\vdash\bot$, we have $\Gamma_p\cup\{\varphi\} \not\vdash_{\Delta_p} \bot$.
Let $q:=(\Delta_p,\Gamma_p\cup\{\varphi\})$ and we have $q\in P$ and $q\geq p$. 
		Since $\varphi$ is atomic, $\varphi\in \Gamma_q\cap\Sen_b(\Delta_q)=f(q)$.
		\item Suppose $p \vdash t_1\stackrel{\act_1\comp\act_2}\Longrightarrow t_2$.
		Let $s$ be the sort of $t_1$ and $t_2$.
		Recall that $C_p$ is the finite set of all Henkin constants from $\Delta_p$.
		Let $c$ be a constant of sort $s$ from $C\setminus C_p$ which does not occur in $t_1$ or $t_2$.
		Let $q:=(\Delta_p[c],\Gamma_p\cup\{t_1\stackrel{\act_1}\Longrightarrow c,c\stackrel{\act_2}\Longrightarrow t_2\})$.
		Notice that $C_q$ is finite. We show that $q\not\vdash\bot$:
		\begin{itemize}
			\item[] 
			Suppose towards a contradiction that $\Gamma_p\cup\{t_1\stackrel{\act_1}\Longrightarrow c,c\stackrel{\act_2}\Longrightarrow t_2\} \vdash_{\Delta_p[c]} \bot$.
			Since $\Gamma_p\vdash_{\Delta_p} t_1\stackrel{\act_1\comp\act_2}\Longrightarrow t_2$, by $(Comp_E)$, $\Gamma_p\vdash_{\Delta_p} \bot$, which is a contradiction.
		\end{itemize}
		Therefore, $q\in P$, $q\geq p$, $q\vdash t_1\stackrel{\act_1}\Longrightarrow c$ and $q\vdash c\stackrel{\act_2}\Longrightarrow t_2$.
		\item Suppose $p \vdash t_1\stackrel{\act_1\cup\act_2}\Longrightarrow t_2$. 
		Let $q_i:=(\Delta_p,\Gamma_p\cup\{t_1\stackrel{\act_i}\Longrightarrow t_2\})$ for all $i\in\{1,2\}$. 
		We show that either $q_1$ or $q_2$ belongs to $P$.
\begin{itemize}
\item [] Suppose towards a contradiction that $\Gamma_p\cup\{t_1\stackrel{\act_i}\Longrightarrow t_2\} \vdash_{\Delta_p} \bot$, for all $i\in\{1,2\}$. 
Since $\Gamma_p\vdash_{\Delta_p} t_1\stackrel{\act_1\cup\act_2}\Longrightarrow t_2$, by $(Union_E)$, we have $\Gamma_p\vdash_{\Delta_p} \bot$, which is a contradiction.
\end{itemize}
		Therefore, there exists $i\in\{1,2\}$ such that $q_i\in P$, $q_i\geq p$, and $q_i\vdash t_1\stackrel{\act_i}\Longrightarrow t_2$.
		\item Suppose $p \vdash t_1\stackrel{\act^*}\Longrightarrow t_2$. 
		Let $q_n\coloneqq (\Delta_p,\Gamma_p\cup\{t_1\stackrel{\act^n}\Longrightarrow t_2\})$ for all $n\in\omega$.  
		We show that $q_n\in P$ for some natural number $n\in\omega$.
		\begin{itemize}
			\item[] Suppose towards a contradiction that $\Gamma_p\cup\{t_1\stackrel{\act^n}\Longrightarrow t_2\} \vdash_{\Delta_p} \bot$ for all natural numbers $n\in\omega$. 
			Since $\Gamma_p \vdash_{\Delta_p} t_1\stackrel{\act^*}\Longrightarrow t_2$, by $(Star_E)$, we have $\Gamma_p \vdash_{\Delta_p} \bot$, which is a contradiction.
		\end{itemize}
		Therefore, there exists $n\in\omega$ such that $q_n\in P$, $q_n\geq p$ and $q_n\vdash t_1\stackrel{\act^n}\Longrightarrow t_2$.
		\item Suppose $p \vdash \vee\Phi$.
		Let $q_\phi\coloneqq (\Delta_p,\Gamma_p\cup\{\phi\})$ for all $\phi\in\Phi$. 
		We show that $q_\phi\in P$ for some $\phi\in\Phi$.
		\begin{itemize}
			\item[] Suppose towards a contradiction that $\Gamma_p\cup\{\phi\} \vdash_{\Delta_p} \bot$ for all $\phi\in\Phi$.
			Since $\Gamma_p \vdash_{\Delta_p} \vee\Phi$, by $(Disj_E)$, we have $\Gamma_p \vdash_{\Delta_p} \bot$, which is a contradiction.
		\end{itemize}
		Therefore, there exists $\phi\in\Phi$ such that $q_\phi\in P$, $q_\phi\geq p$, and $q_\phi\vdash \phi$.
		\item Suppose $p \vdash \Exists{X}\phi$.
		Notice that $p'\coloneqq (\Delta_p,\Gamma_p\cup\{\Exists{X}\phi\})\in P$.
		Let $\theta:X\to C\setminus C_{p'}$	be an injective mapping.
		We show that $q\coloneqq (\Delta_p[\theta(X)], \Gamma_p\cup\{\theta(\phi)\})$ belongs to $P$.
		\begin{itemize}
			\item[] Suppose towards a contradiction that $\Gamma_p\cup\{\theta(\phi)\}\vdash_{\Delta_q}\bot$.
			Let $\chi:\Delta_q\to \Delta_p[X]$ be the signature morphism which is the identity on $\Delta_p$ and maps each $\theta(x)$ to $x$.
			Translate $\Gamma_p\cup\{\theta(\phi)\} \vdash_{\Delta_q}\bot$ along $\chi$ to obtain $\Gamma_p\cup\{\phi\} \vdash_{\Delta_p[X]}\bot$.
			By $(Quant_I)$, we get $\Gamma_p\cup\{\Exists{X}\phi\} \vdash_{\Delta_p}\bot$, which is a contradiction with  $\Gamma_p \vdash_{\Delta_p} \Exists{X}\phi$.
		\end{itemize}
		Therefore, $q\in P$, $q\geq p$ and $q\vdash\theta(\phi)$.
	\end{enumerate}
\end{proof}
\begin{proof}[Proof of Theorem~\ref{th:sfp}]
We proceed by induction on the structure of the sentence $\phi$.
	\begin{proofcases}
		\item[$\varphi$ is atomic]\
		\begin{proofcases}
			\item[$\Rightarrow$] Assume $p\Vdash^{w} \varphi$.
			\begin{proofsteps}{17em}
				let $q\geq p$ be a condition
				&
				\\
				\label{thVwD-atm-imp-2}
				$r\Vdash \varphi$ for some $r\geq q$
				& by Lemma~\ref{lemma:fp}~(\ref{fp-1}), since $p\Vdash^{w} \varphi$
				\\
				$r\vdash \varphi$
				& since $\varphi\in f(r)\subseteq \Gamma_r$ 
			\end{proofsteps}
			Since for all $q\geq p$ there is $r\geq q$ such that $r\vdash \varphi$, by Lemma~\ref{lemma:vD=vD^w}, $p\vdash \varphi$.
			\item[$\Leftarrow$] Assume $p\vdash \varphi$.
			\begin{proofsteps}{17em}
				let $q\geq p$ be a condition
				&
				\\
				$q\vdash \varphi$
				& by assumption
				\\
				$\varphi\in f(r)$ for some $r\geq q$
				& by Lemma~\ref{lemma:up} 
				\\
				$r\Vdash \varphi$
				& by the definition of forcing relation
				\\
				$p\Vdash^w \varphi$
				& since $q\geq p$ was arbitrarily chosen and $r\geq q$
			\end{proofsteps}
		\end{proofcases}
		\item[$t_1\stackrel{\act_1\comp\act_2}\Longrightarrow  t_2$]\
		\begin{proofcases}
			\item[$\Rightarrow$] Assume $p\Vdash^{w} t_1\stackrel{\act_1\comp\act_2}\Longrightarrow  t_2$.
			\begin{proofsteps}{17em}
				let $q\geq p$ be a condition
				&
				\\
				$r\Vdash t_1\stackrel{\act_1\comp\act_2}\Longrightarrow  t_2$ for some $r\geq q$
				& by Lemma~\ref{lemma:fp}~(\ref{fp-1}), since $p\Vdash^{w} t_1\stackrel{\act_1\comp\act_2}\Longrightarrow  t_2$
				\\
				$r\Vdash t_1\stackrel{\act_1}\Longrightarrow  t$ and $r\Vdash t\stackrel{\act_2}\Longrightarrow  t_2$ \newline
				for some $t\in T_{\Delta_r}$
				& by the definition of forcing relation
				\\
				$r\Vdash^w t_1\stackrel{\act_1}\Longrightarrow  t$ and $r\Vdash^w t\stackrel{\act_2}\Longrightarrow  t_2$
				& by Lemma~\ref{lemma:fp}~(\ref{fp-3})
				\\
				$r\vdash t_1\stackrel{\act_1}\Longrightarrow  t$ and $r\vdash t\stackrel{\act_2}\Longrightarrow  t_2$
				& by induction hypothesis
				\\
				$r\vdash t_1\stackrel{\act_1\comp\act_2}\Longrightarrow  t_2$
				& by $(Comp_I)$
			\end{proofsteps}
			Since for all $q\geq p$ there is $r\geq q$ such that $r\vdash t_1\stackrel{\act_1\comp\act_2}\Longrightarrow  t_2$, by Lemma~\ref{lemma:vD=vD^w}, $p\vdash t_1\stackrel{\act_1\comp\act_2}\Longrightarrow  t_2$.
			\item[$\Leftarrow$] Assume $p\vdash t_1\stackrel{\act_1\comp\act_2}\Longrightarrow  t_2$.
			\begin{proofsteps}{15em}
				let $q\geq p$ be a condition
				&
				\\
				$q\vdash t_1\stackrel{\act_1\comp\act_2}\Longrightarrow  t_2$
				& since $p\vdash t_1\stackrel{\act_1\comp\act_2}\Longrightarrow  t_2$
				\\
				$r'\vdash t_1\stackrel{\act_1}\Longrightarrow  t$ and $r'\vdash t\stackrel{\act_2}\Longrightarrow  t_2$ 
				\newline for some $r'\geq q$ and some $t\in T_{\Delta_{r'}}$
				& by Lemma~\ref{lemma:up}~(\ref{lemma:up2})
				\\
				\label{thVwD-comp-if-4}
				$r'\Vdash^w t_1\stackrel{\act_1}\Longrightarrow  t$ and $r'\Vdash^w t\stackrel{\act_2}\Longrightarrow  t_2$
				& by induction hypothesis
				\\
				\label{thVwD-comp-if-5}
				$r''\Vdash t_1\stackrel{\act_1}\Longrightarrow  t$ for some $r''\geq r'$
				& by Lemma~\ref{lemma:fp}~(\ref{fp-1}), since $r'\geq r'$ and $r'\Vdash^w t_1\stackrel{\act_1}\Longrightarrow  t$
				\\
				\label{thVwD-comp-if-6}
				$r\Vdash t\stackrel{\act_2}\Longrightarrow  t_2$ for some $r\geq r''$
				& by Lemma~\ref{lemma:fp}~(\ref{fp-1}), since $r''\geq r'$ and $r'\Vdash^w t\stackrel{\act_2}\Longrightarrow  t_2$
				\\
				\label{thVwD-comp-if-7}
				$r\Vdash t_1\stackrel{\act_1}\Longrightarrow  t$
				& by Lemma~\ref{lemma:fp}~(\ref{fp-2}) since $r\geq r''$ and $r''\Vdash t_1\stackrel{\act_1}\Longrightarrow  t$
				\\
				$r\Vdash t_1\stackrel{\act_1\comp\act_2}\Longrightarrow  t_2$
				& from \ref{thVwD-comp-if-6} and \ref{thVwD-comp-if-7} 
			\end{proofsteps}
		\end{proofcases}
		\item[$t_1\stackrel{\act_1\cup\act_2}\Longrightarrow  t_2$]\
		\begin{proofcases}
			\item[$\Rightarrow$] Assume $p\Vdash^{w} t_1\stackrel{\act_1\cup\act_2}\Longrightarrow  t_2$.
			\begin{proofsteps}{17em}
				let $q\geq p$ be a condition
				&
				\\
				$r\Vdash t_1\stackrel{\act_1\cup\act_2}\Longrightarrow  t_2$ for some $r\geq q$
				& by Lemma~\ref{lemma:fp}~(\ref{fp-1}), since $p\Vdash^{w} t_1\stackrel{\act_1\cup\act_2}\Longrightarrow  t_2$
				\\
				$r\Vdash t_1\stackrel{\act_1}\Longrightarrow  t_2$ or $r\Vdash t_1\stackrel{\act_2}\Longrightarrow  t_2$
				& by the definition of forcing relation
				\\
				$r\Vdash^w t_1\stackrel{\act_1}\Longrightarrow  t_2$ or $r\Vdash^w t_1\stackrel{\act_2}\Longrightarrow  t_2$
				& by Lemma~\ref{lemma:fp}~(\ref{fp-3})
				\\
				$r\vdash t_1\stackrel{\act_1}\Longrightarrow  t_2$ or $r\vdash t_1\stackrel{\act_2}\Longrightarrow  t_2$
				& by induction hypothesis
				\\
				$r\vdash t_1\stackrel{\act_1\cup\act_2}\Longrightarrow  t_2$
				& by $(Union_I)$
			\end{proofsteps}
			Since for all $q\geq p$ there is $r\geq q$ such that $r\vdash t_1\stackrel{\act_1\cup\act_2}\Longrightarrow  t_2$, by Lemma~\ref{lemma:vD=vD^w}, $p\vdash t_1\stackrel{\act_1\cup\act_2}\Longrightarrow  t_2$.
			\item[$\Leftarrow$] Assume $p\vdash t_1\stackrel{\act_1\cup\act_2}\Longrightarrow  t_2$.
			\begin{proofsteps}{20em}
				let $q\geq p$ be a condition
				&
				\\
				$q\vdash t_1\stackrel{\act_1\cup\act_2}\Longrightarrow  t_2$
				& since $p\vdash t_1\stackrel{\act_1\cup\act_2}\Longrightarrow  t_2$
				\\
				$r'\vdash t_1\stackrel{\act_1}\Longrightarrow  t_2$ or $r'\vdash t_1\stackrel{\act_2}\Longrightarrow  t_2$ for some $r'\geq q$
				& by Lemma~\ref{lemma:up}~(\ref{lemma:up3})
				\\
				$r'\Vdash^w t_1\stackrel{\act_1}\Longrightarrow  t_2$ or $r'\Vdash^w t_1\stackrel{\act_2}\Longrightarrow  t_2$
				& by induction hypothesis
				\\
				$r\Vdash t_1\stackrel{\act_1}\Longrightarrow  t_2$ or $r\Vdash t_1\stackrel{\act_2}\Longrightarrow  t_2$ for some $r\geq r'$
				& by Lemma~\ref{lemma:fp}~(\ref{fp-1})
				\\
				$r\Vdash t_1\stackrel{\act_1\cup\act_2}\Longrightarrow  t_2$
				& by the definition of forcing relation
				\\
				$p\Vdash^w t_1\stackrel{\act_1\cup\act_2}\Longrightarrow  t_2$
				& since $q\geq p$ was arbitrarily chosen and $r\geq q$
			\end{proofsteps}
		\end{proofcases}
		\item[$t_1\stackrel{\act^*}\Longrightarrow  t_2$]\
		
		\begin{proofcases}
			\item[$\Rightarrow$] Assume $p\Vdash^{w} t_1\stackrel{\act^*}\Longrightarrow  t_2$.
			\begin{proofsteps}{17em}
				let $q\geq p$ be a condition
				&
				\\
				$r\Vdash t_1\stackrel{\act^*}\Longrightarrow  t_2$ for some $r\geq q$
				& by Lemma~\ref{lemma:fp}~(\ref{fp-1}), since $p\Vdash^{w} t_1\stackrel{\act^*}\Longrightarrow  t_2$
				\\
				$r\Vdash t_1\stackrel{\act^n}\Longrightarrow  t_2$ for some $n\in \omega$
				& by the definition of forcing relation
				\\
				$r\Vdash^w t_1\stackrel{\act^n}\Longrightarrow  t_2$
				& by Lemma~\ref{lemma:fp}~(\ref{fp-3})
				\\
				$r\vdash t_1\stackrel{\act^n}\Longrightarrow  t_2$
				& by induction hypothesis
				\\
				$r\vdash t_1\stackrel{\act^*}\Longrightarrow  t_2$
				& by $(Star_I)$
			\end{proofsteps}
			Since for all $q\geq p$ there is $r\geq q$ such that $r\vdash t_1\stackrel{\act^*}\Longrightarrow  t_2$, by Lemma~\ref{lemma:vD=vD^w}, $p\vdash t_1\stackrel{\act^*}\Longrightarrow  t_2$.
			\item[$\Leftarrow$] Assume $p\vdash t_1\stackrel{\act^*}\Longrightarrow  t_2$.
			\begin{proofsteps}{18em}
				let $q\geq p$ be a condition
				&
				\\
				$q\vdash t_1\stackrel{\act^*}\Longrightarrow  t_2$
				& since $p\vdash t_1\stackrel{\act^*}\Longrightarrow  t_2$
				\\
				$r'\vdash t_1\stackrel{\act^n}\Longrightarrow  t_2$ for some $r'\geq q$ and $n\in \omega$
				& by Lemma~\ref{lemma:up}~(\ref{lemma:up4})
				\\
				$r'\Vdash^w t_1\stackrel{\act^n}\Longrightarrow  t_2$
				& by induction hypothesis
				\\
				$r\Vdash t_1\stackrel{\act^n}\Longrightarrow  t_2$ for some $r\geq r'$
				& by Lemma~\ref{lemma:fp}~(\ref{fp-1})
				\\
				$r\Vdash t_1\stackrel{\act^*}\Longrightarrow  t_2$
				& by the definition of forcing relation
				\\
				$p\Vdash^w t_1\stackrel{\act^*}\Longrightarrow  t_2$
				& since $q\geq p$ was arbitrarily chosen
			\end{proofsteps}
		\end{proofcases}
		\item[$\neg\phi$]\
		\begin{proofcases}
			\item[$\Rightarrow$] Assume $p\Vdash^{w} \neg\phi$.
			\begin{proofsteps}{17em}
				let $q\geq p$ be a condition 
				& \\
				$p\Vdash \neg\neg\neg\phi$
				& since $p\Vdash^{w} \neg\phi$
				\\
				$q\not\Vdash \neg\neg\phi$ 
				& by the definition of forcing relation
				\\
				$q\not\Vdash^w \phi$
				& by the definition of $\Vdash^w$
				\\
				$q\not\vdash \phi$
				& by induction hypothesis
				\\
				$(\Delta_q,\Gamma_q\cup\{\neg\phi\})\in P$
				& since $\Gamma_q\not\vdash_{\Delta_q} \phi$
				\\
				$r\vdash \neg\phi$, where $r:=(\Delta_q,\Gamma_q\cup\{\neg\phi\})$
			\end{proofsteps}
			Since for all $q\geq p$ there is $r\geq q$ such that $r\vdash \neg\phi$, by Lemma~\ref{lemma:vD=vD^w}, $p\vdash \neg\phi$.
			\item[$\Leftarrow$] Assume $p\vdash \neg\phi$.
			\begin{proofsteps}{17em}
				$q\vdash  \neg\phi$ for all $q\geq p$ & 
				since $p\vdash \neg\phi$ \\
				$q\not\vdash \phi$ for all $q\geq p$
				& since $q$ is consistent
				\\
				$q\not\Vdash^w \phi$ for all $q\geq p$
				& by induction hypothesis
				\\
				$q\not\Vdash \neg\neg\phi$ for all $q\geq p$
				& by the definition of $\Vdash^w$
				\\
				$p\Vdash \neg\neg\neg\phi$
				& by the definition of forcing relation
				\\
				$p\Vdash^{w} \neg\phi$
				& by the definition of $\Vdash^w$
			\end{proofsteps}
		\end{proofcases}
		\item[$\vee\Phi$]\
		\begin{proofcases}
			\item[$\Rightarrow$] Assume $p\Vdash^{w} \vee\Phi$.
			\begin{proofsteps}{17em}
				let $q\geq p$ be a condition
				&
				\\
				$r\Vdash \vee\Phi$ for some $r\geq q$
				& by Lemma~\ref{lemma:fp}~(\ref{fp-1}), since $p\Vdash^{w} \vee\Phi$
				\\
				$r\Vdash \phi$ for some $\phi\in \Phi$
				& by the definition of forcing relation
				\\
				$r\Vdash^w \phi$
				& by Lemma~\ref{lemma:fp}~(\ref{fp-3})
				\\
				$r\vdash \phi$
				& by induction hypothesis
				\\
				$r\vdash \vee\Phi$
				& by $(Disj_I)$
			\end{proofsteps}
			Since for all $q\geq p$ there is $r\geq q$ such that $r\vdash \vee\Phi$, by Lemma~\ref{lemma:vD=vD^w}, $p\vdash \vee\Phi$.
			\item[$\Leftarrow$] Assume $p\vdash \vee\Phi$.
			\begin{proofsteps}{18em}
				let $q\geq p$ be a condition
				&
				\\
				$q\vdash \vee\Phi$
				& since  $p\vdash \vee\Phi$
				\\
				$r'\vdash \phi$ for some $r'\geq q$ and some $\phi\in \Phi$
				& by Lemma~\ref{lemma:up}~(\ref{lemma:up5})
				\\
				$r'\Vdash^w \phi$
				& by induction hypothesis
				\\
				$r\Vdash \phi$ for some $r\geq r'$
				& by Lemma~\ref{lemma:fp}~(\ref{fp-1})
				\\
				$r\Vdash \vee\Phi$
				& by the definition of forcing relation
				\\
				$p\Vdash^w \vee\Phi$
				& since $q\geq p$ was arbitrarily chosen and $r\geq q$
			\end{proofsteps}
		\end{proofcases}
		\item[$\Exists{X}\phi$]\
		\begin{proofcases}
			\item[$\Rightarrow$] Assume $p\Vdash^{w} \Exists{X}\phi$.
			\begin{proofsteps}{17em}
				let $q\geq p$ be a condition
				&
				\\
				$r\Vdash \Exists{X}\phi$ for some $r\geq q$
				& by Lemma~\ref{lemma:fp}~(\ref{fp-1}), since $p\Vdash^{w} \Exists{X}\phi$
				\\
				$r\Vdash \theta(\phi)$ for some $\theta:X\to T_{\Delta_r}$
				& by the definition of forcing relation
				\\
				$r\Vdash^w \theta(\phi)$
				& by Lemma~\ref{lemma:fp}~(\ref{fp-3})
				\\
				$r\vdash \theta(\phi)$
				& by induction hypothesis
				\\
				$r\vdash \Exists{X}\phi$
				& by $(Subst)$
			\end{proofsteps}
			Since for all $q\geq p$ there is $r\geq q$ such that $r\vdash \Exists{X}\phi$, by Lemma~\ref{lemma:vD=vD^w}, $p\vdash \Exists{X}\phi$.			
			\item[$\Leftarrow$] Assume $p\vdash \Exists{X}\phi$.
			\begin{proofsteps}{19em}
				let $q\geq p$ be a condition
				&
				\\
				$q\vdash \Exists{X}\phi$
				& since $p\vdash \Exists{X}\phi$
				\\
				$r'\vdash \theta(\phi)$ for some $r'\geq q$ and $\theta:X\to T_{\Delta_{r'}}$
				& by Lemma~\ref{lemma:up}~(\ref{lemma:up6})
				\\
				$r'\Vdash^w \theta(\phi)$
				& by induction hypothesis
				\\
				$r\Vdash \theta(\phi)$ for some $r\geq r'$
				& by Lemma~\ref{lemma:fp}~(\ref{fp-1})
				\\
				$r\Vdash \Exists{X}\phi$
				& by the definition of forcing relation
				\\
				$p\Vdash^w \Exists{X}\phi$
				& since $q\geq p$ was arbitrarily chosen $r\geq q$
			\end{proofsteps}
		\end{proofcases}
	\end{proofcases}
\end{proof}
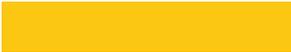
\end{document}